%% file: Main.tex
\documentclass[preprint,tightenlines,superscriptaddress,amsmath,amssymb,aps,prb,floatfix]{revtex4-2}

\usepackage{graphicx}
\usepackage[dvipsnames]{xcolor}
\usepackage[hidelinks]{hyperref}
\usepackage{siunitx}
\usepackage{bm}
\usepackage{tikz}

\makeatletter
\newsavebox{\@brx}
\newcommand{\llangle}[1][]{\savebox{\@brx}{\(\m@th{#1\langle}\)}%
  \mathopen{\copy\@brx\kern-0.5\wd\@brx\usebox{\@brx}}}
\newcommand{\rrangle}[1][]{\savebox{\@brx}{\(\m@th{#1\rangle}\)}%
  \mathclose{\copy\@brx\kern-0.5\wd\@brx\usebox{\@brx}}}
\makeatother

\renewcommand{\figurename}{Fig.}
\renewcommand{\tablename}{Table}

\newcounter{masterfig}
\newcounter{mastertab}
\newcounter{mastereq}
\makeatletter
\@addtoreset{figure}{masterfig}
\@addtoreset{table}{mastertab}
\@addtoreset{equation}{mastereq}
\makeatother

\begin{document}
\title{Evidence of Coulomb liquid phase in few-electron droplets}

\author{Jashwanth Shaju}
    \altaffiliation{These authors contributed equally to this work.}
    \affiliation{Universit\'e Grenoble Alpes, CNRS, Grenoble INP, Institut N\'eel, F-38000 Grenoble, France}

\author{Elina Pavlovska}
    \altaffiliation{These authors contributed equally to this work.}
    \affiliation{Department of Physics, University of Latvia, Riga, LV-1004, Latvia}
    
\author{Ralfs Suba}
    \affiliation{Department of Physics, University of Latvia, Riga, LV-1004, Latvia}
    
\author{Junliang Wang}
    \affiliation{Universit\'e Grenoble Alpes, CNRS, Grenoble INP, Institut N\'eel, F-38000 Grenoble, France}

\author{Seddik Ouacel}
    \affiliation{Universit\'e Grenoble Alpes, CNRS, Grenoble INP, Institut N\'eel, F-38000 Grenoble, France}

\author{Thomas Vasselon}
    \affiliation{Universit\'e Grenoble Alpes, CNRS, Grenoble INP, Institut N\'eel, F-38000 Grenoble, France}

\author{Matteo Aluffi}
    \affiliation{Universit\'e Grenoble Alpes, CNRS, Grenoble INP, Institut N\'eel, F-38000 Grenoble, France}

\author{Lucas Mazzella}
    \affiliation{Universit\'e Grenoble Alpes, CNRS, Grenoble INP, Institut N\'eel, F-38000 Grenoble, France}

\author{Cl\'ement Geffroy}
    \affiliation{Universit\'e Grenoble Alpes, CNRS, Grenoble INP, Institut N\'eel, F-38000 Grenoble, France}

\author{Arne Ludwig}
    \affiliation{Lehrstuhl f\"ur Angewandte Festk\"orperphysik, Ruhr-Universit\"at Bochum, Universit\"atsstraße 150, D-44780 Bochum, Germany}
    
\author{Andreas D. Wieck}
    \affiliation{Lehrstuhl f\"ur Angewandte Festk\"orperphysik, Ruhr-Universit\"at Bochum, Universit\"atsstraße 150, D-44780 Bochum, Germany}
 
\author{Matias Urdampilleta}
    \affiliation{Universit\'e Grenoble Alpes, CNRS, Grenoble INP, Institut N\'eel, F-38000 Grenoble, France}
    
\author{Christopher B\"auerle}
    \affiliation{Universit\'e Grenoble Alpes, CNRS, Grenoble INP, Institut N\'eel, F-38000 Grenoble, France}

\author{Vyacheslavs Kashcheyevs}
    \altaffiliation{corresponding author: \href{mailto: slava@latnet.lv}{slava@latnet.lv}}
    \affiliation{Department of Physics, University of Latvia, Riga, LV-1004, Latvia}

\author{Hermann Sellier}
    \altaffiliation{corresponding author: \href{mailto: hermann.sellier@neel.cnrs.fr}{hermann.sellier@neel.cnrs.fr}}
    \affiliation{Universit\'e Grenoble Alpes, CNRS, Grenoble INP, Institut N\'eel, F-38000 Grenoble, France}

\maketitle


Emergence of universal collective behaviour from interactions within a sufficiently large group of elementary constituents is a fundamental scientific paradigm~\cite{anderson1972more}. 
In physics, correlations in fluctuating microscopic observables can provide key information about collective states of matter such as deconfined quark-gluon plasma in heavy-ion collisions~\cite{gupta2011scale} or expanding quantum degenerate gases~\cite{demarco1999onset,bloch2012quantum}.
Mesoscopic colliders, through shot-noise measurements, have provided smoking-gun evidence on the nature of exotic electronic excitations such as fractional charges~\cite{saminadayar1997observation,picotto1997direct}, levitons~\cite{dubois2013minimal} and anyon statistics~\cite{bartolomei2020fractional}. 
Yet, bridging the gap between two-particle collisions and the emergence of collectivity~\cite{grosse2024decade} as the number of interacting particles increases~\cite{wenz2013few} remains a challenging task at the microscopic level.
Here we demonstrate all-body correlations in the partitioning of electron droplets containing up to $N=5$ electrons, driven by a moving potential well through a Y-junction in a semiconductor device.
Analyzing the partitioning data using high-order multivariate cumulants and finite-size scaling towards the thermodynamic limit reveals distinctive fingerprints of a strongly-correlated Coulomb liquid.
These fingerprints agree well with a universal limit where the partitioning of a droplet is predicted by a single collective variable.
Our electron-droplet collider provides critical insight into the interplay of confinement and interaction effects in small electron systems and highlights a new way to study engineered states of matter.

\newpage
\noindent
\textbf{INTRODUCTION}

\noindent
Breaking-up matter into pieces and studying the statistics of fragments is one of the basic epistemic strategies in physics. 
Arguably the most exquisite pursuit of this strategy is the success of high-energy particle colliders in discovering and quantifying the fundamental types of matter within the Standard Model of elementary particles. 
The quantum chromodynamics (QCD) sector of the latter is particularly challenging as the fundamental particles of QCD --- quarks and gluons --- at low energies condense into strongly correlated ``droplets’’ (hadrons) due to the phenomenon of color confinement.  
A deconfinement phase transition from a liquid of hadrons into a quark-gluon plasma~\cite{harris1996search} has been extensively studied in relativistic heavy-ion collisions, where pairs of colliding nuclei create a fireball of high baryonic density and temperature. 
In particular, measurements of high-order cumulants in the number of particles (collision multiplicity), produced as the fireball quenches~\cite{adamczyk2014energy}, can be used~\cite{gupta2011scale} to pinpoint the critical point in the QCD phase diagram~\cite{rajagopal1993static,harris2024qgp}. 
Fluctuations in multiplicity and correlations of the detected transverse momenta carry rich information about the collective dynamics of QCD matter that continues to be a vibrant area of research at the interface of heavy-ion and high-energy physics~\cite{grosse2024decade}.

In solid-state nanoelectronic circuits, charged quasi-particles can be launched with on-demand single-electron sources and guided to a small interaction area such as a quantum point contact~\cite{dubois2013minimal, bocquillon2013coherence} or an energy barrier~\cite{wang2023coulomb, ubbelohde2023two, fletcher2023time}, creating a collider analogue for electronic matter. 
Second-order correlations in steady-state and on-demand collisions have provided an essential tool to decode partitioning noise of composite particles~\cite{saminadayar1997observation,picotto1997direct}, fermionic~\cite{dubois2013minimal,bocquillon2013coherence} and anyonic~\cite{bartolomei2020fractional} exchange statistics, and two-particle Coulomb interactions~\cite{wang2023coulomb, ubbelohde2023two, fletcher2023time}. 
In previous research, temporal electronic correlations in nanostructures have been extensively studied~\cite{levitov1996electron,lu2003real, fujisawa2006bidirectional, gustavsson2006counting, ubbelohde2012measurement,  bylander2005current, bomze2005measurement}.
These experiments primarily investigated the Coulomb interaction between two neighbouring electrons as they traverse a quantum point contact, quantum dot, or tunnel junction. 

Higher-order ($k>2$) correlations in current fluctuations~\cite{reulet2003environmental, forgues2013noise} and electron counting statistics~\cite{gershon2008detection,flindt2009universal,ubbelohde2012measurement} have been recognised as important signatures of Coulomb interactions. 
Yet evidence for the corresponding collective behaviour in on-chip transport has been difficult to interpret~\cite{kambly2011factorial} due to limited control over the number $N$ of interacting particles and the dominating randomness of tunneling times. 
Investigating the gap between few-particle correlations and the thermodynamic limit \cite{wenz2013few} for two-dimensional electron systems is motivated by their rich phase diagram as function of electron density, temperature and magnetic field, including strongly-correlated Coulomb liquid, Wigner crystals and quantum Hall phases~\cite{monarkha2013two}.

Here, by drawing an analogy with relativistic ion collisions, we investigate partitioning of a small electron-plasma droplet containing a precise number $N$ of electrons.
Analysing the partitioning data of our synthesised electron droplet using multivariate cumulants enables us to identify its corresponding strongly-correlated state of electronic matter.

\begin{figure*}[t]
    \includegraphics[scale=1]{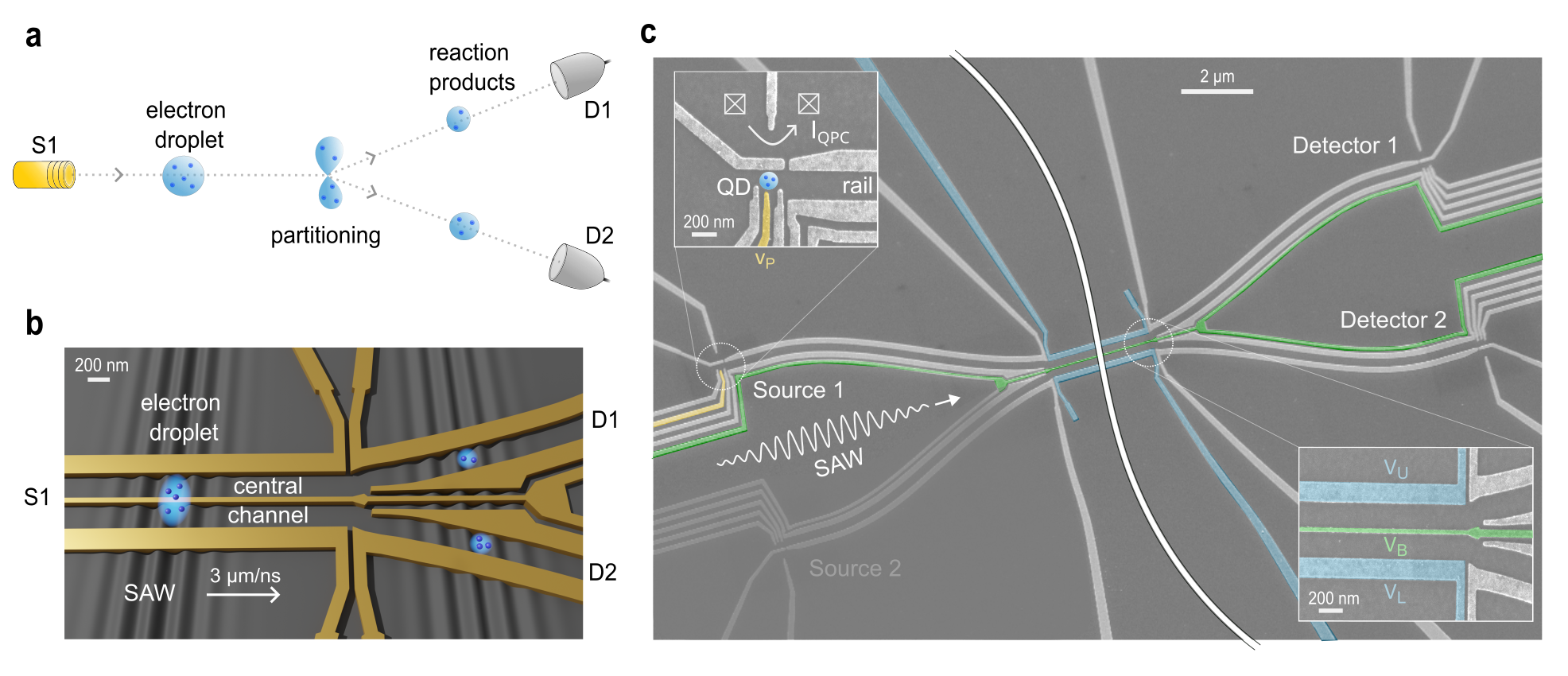}
    \caption{\textbf{Partitioning of an electron droplet.} 
    \textbf{a}, Schematic of the experiment. 
    An electron source (S1) delivers a few-electron droplet which is split in-flight at a Y-junction. 
    The output of the partitioning is analysed by two single-shot detectors (D1 and D2).
    \textbf{b}, Schematic of the electron droplet transport inside the selected potential minimum of a surface acoustic wave (SAW). 
    Electrostatic gates (yellow) are used to guide the electron droplet and create a Y-junction. 
    \textbf{c}, Scanning electron microscope image of the device showing the metallic surface gates (light grey). 
    The electron source (S1) consists of a quantum dot (shown in the top left inset) coupled to a quantum point contact (QPC) for charge sensing. 
    The plunger gate (yellow) is employed to inject a precise number of electrons into a single SAW minimum. 
    A second electron source (S2) is connected to the central channel to inject more electrons. 
    The Y-junction at the end of the central channel (see bottom inset) enables partitioning of the electron droplet.
    }
    \label{figure1}
\end{figure*}

\vspace{3mm}
\noindent
\textbf{ON-CHIP MULTI-ELECTRON COLLIDER}

\noindent
We have implemented the partitioning of a charge droplet of interacting electrons using a Y-junction in a GaAs semiconductor heterostructure, as illustrated in Fig.~\ref{figure1}.
Two single-electron sources and two single-shot detectors are made of gate-defined quantum dots (QD) paired with nearby quantum point contacts (QPC) used as charge sensors. 
By recording the discrete jumps in the QPC current $I_{\rm QPC}$, the precise electron number within the QD is measured. 
Several pairs of parallel electrodes define depleted quasi-one-dimensional transport rails, guiding the electron droplet from the sources to the detectors. 
A \SI{40}{\micro\meter}-long central channel is employed to control the droplet properties before partitioning.
This channel includes a narrow \SI{30}{\nano\meter}-wide barrier gate that enables precise tuning of the confining potential in the direction perpendicular to the rail.
At the end of the channel, a Y-junction splits the electron droplet into two parts and directs the ``reaction products'' towards the detectors. 
The counting statistics is accumulated into the probabilities $P_{(N-n,\,n)}$, where $n$ and $N-n$ are the numbers of electrons measured, after each single-shot partitioning, in detectors D1 and D2, respectively.

In our experiment, the electron droplet is transported within a single piezoelectric potential minimum of a surface acoustic wave (SAW)~\cite{hermelin2011electrons,mcneil2011demand,takada2019sound}.
An interdigital transducer, positioned \SI{1.5}{\milli\meter} away, generates a \SI{180}{\micro\meter}-long SAW train.
By applying a voltage pulse $V_\text{P}$ on the plunger gate of the source quantum dot, with a duration much shorter than the SAW period, a well-defined number of electrons (ranging from 1 to 5) can be loaded into a single minimum of the SAW potential (see Supplementary Note~\ref{supp:transfer}). 
When the SAW propagates across the device, these electrons remain confined in the moving quantum dot~\cite{edlbauer2021flight} which shuttles them along the rails.

For droplets with more than three electrons, the injection from a single source becomes technically challenging, and we instead prepare these droplets using two sources, synchronised to the same SAW minimum. 
The two parts then merge in the central channel where the voltage $V_\text{B}$ on the barrier gate is tuned to ensure that the electrons loose their history and become statistically indistinguishable (see Supplementary Note~\ref{supp:indistinguishability}).

\begin{figure*}[t]
    \includegraphics[scale=0.95]{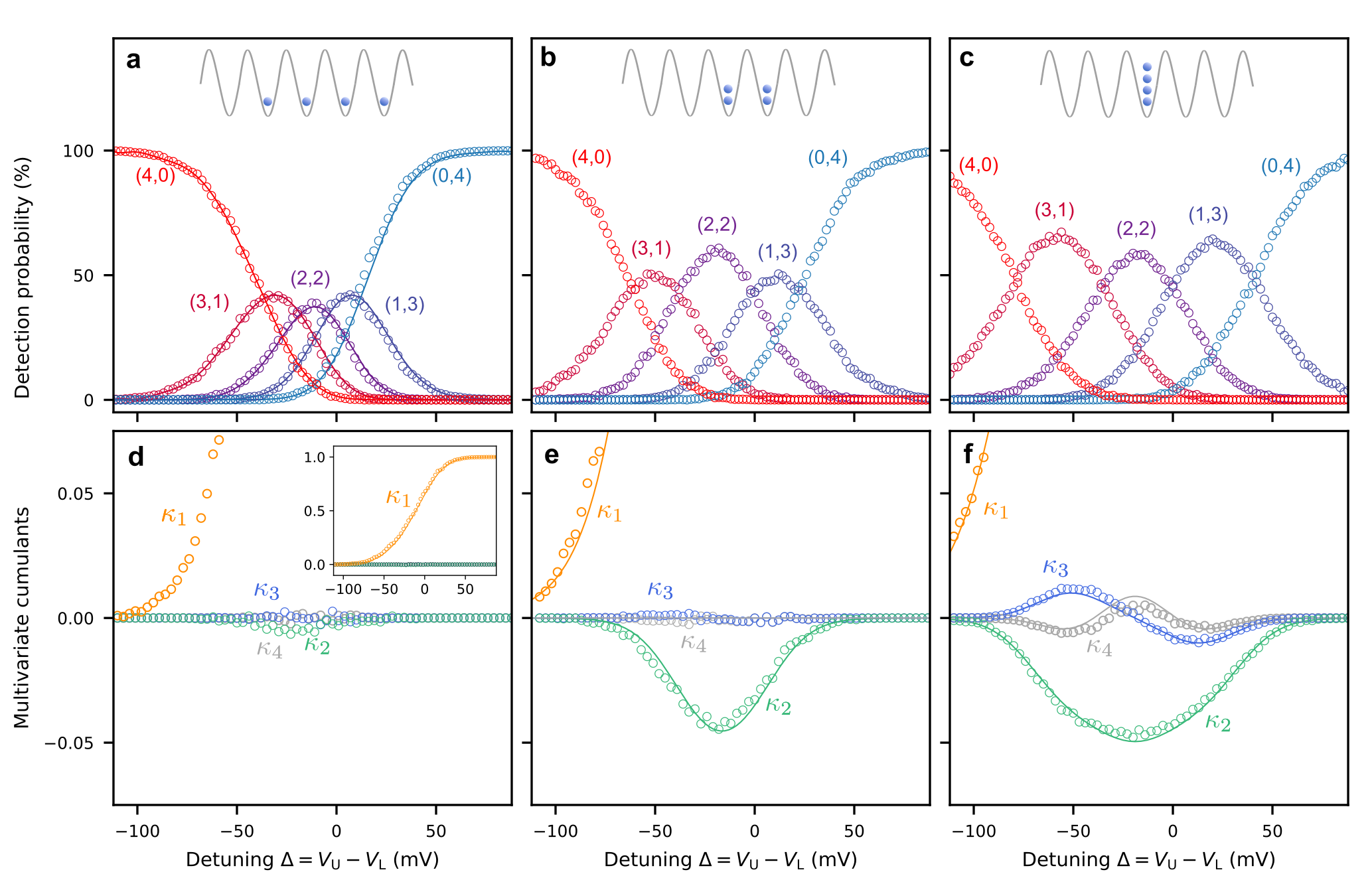}
    \caption{\textbf{Partitioning of an electron droplet containing $N=4$ electrons.} 
    \textbf{a-c}, Detection probabilities $P_{(N-n,\,n)}$ versus side-gate detuning voltage $\Delta$. 
    Each data point is extracted from 3000 single-shot measurements. 
    The labels $(N-n,\,n)$ correspond to the events where $n$ electrons are measured at detector D1 and $N-n$ electrons at detector D2.
    In \textbf{a}, the four electrons are distributed across different SAW minima, as illustrated in the top inset. 
    In \textbf{b}, the four electrons are loaded into adjacent minima, with two electrons in each.
    In \textbf{c}, all four electrons are confined into a single minimum. 
    The solid lines in \textbf{a} are predictions based on independently measured single electron partitioning probabilities (see Supplementary Note~\ref{supp:reconstruction}).
    \textbf{d-f}, Multivariate cumulants $\kappa_1\ldots\kappa_N$ calculated from the measured probabilities shown in \textbf{a-c}.
    The inset in \textbf{d} shows the evolution of $\kappa_1$ across the entire range, and the solid line is the partitioning probability $P_{(0,1)}$ of a single electron.
    In \textbf{e}, two non-equivalent cumulants contribute to $\kappa_2$ (see Methods and Supplementary Note~\ref{supp:2e2epartitioning}).
    Solid lines in \textbf{e} and \textbf{f} are fits using the Ising model of Eq.~\eqref{eq:Hamiltonian}.
    }
    \label{figure2}
\end{figure*}

\vspace{3mm}
\noindent
\textbf{PARTITIONING OF AN ELECTRON DROPLET}

\noindent
We illustrate our ability to control the partitioning in Fig.~\ref{figure2} where the counting statistics $P_{(N-n,n)}$ is shown for $N=4$ as function of the voltage difference $\Delta=V_\text{U}-V_\text{L}$ between the two electrodes defining the central channel. 
This parameter acts as a tuneable impact parameter for the collision with the Y-junction.

The simplest case is when all the electrons are placed in different SAW minima (Fig.~\ref{figure2}a) such that they are prevented from forming a droplet and cannot interact. 
We find that the counting statistics $P_{(N-n,\,n)}$ can be reconstructed from single-electron partitioning data (solid lines) and thus follows a binomial distribution, with electrons scattering at the Y-junction independently of each other. 
Such statistics corresponds to fixed-$N$ samples of a non-interacting electron gas.

To induce correlations, we  group the electrons in two pairs, placed in adjacent SAW minima (Fig.~\ref{figure2}b). 
An increase of the probability $P_{(2,2)}$ can be observed compared to the non-interacting case, indicating antibunching of the two electrons contained in each pair~\cite{wang2023coulomb}.
To obtain a strongly-correlated state, we place all four electrons in the same SAW minimum (Fig.~\ref{figure2}c) and note a similar increase in $P_{(2,2)}$ but the maxima of $P_{(1,3)}$ and $P_{(3,1)}$ now exceed $P_{(2,2)}$. 
While the probabilities in \ref{figure2}b and \ref{figure2}c are qualitatively different, the multi-electron interdependencies are difficult to interpret directly from the counting statistics.

\vspace{3mm}
\noindent
\textbf{MULTIVARIATE CUMULANTS}

\noindent
To interrogate the nature of the many-electron state in the droplet, we aim to characterise its internal correlations and decompose them into irreducible components, known as cumulants~\cite{fisher1932pattern}.
Cumulants are convenient as they capture not only pairwise but also higher-order correlations.
This is crucial for understanding complex many-body systems where strong enough pairwise interactions can lead to correlations of all orders, heralding the emergence of a new collective state.
One possibility is to consider the high-order cumulants $\llangle n^k\rrangle$, or their combinations such as skewness and kurtosis, of the collective variable $n$, as it is measured directly~\cite{gupta2011scale,flindt2009universal}. 
Yet, in this representation, contributions of individual particles are not resolved. 
To elucidate few-electron correlation effects, it is crucial to separate these correlations by order, corresponding to the number of particles involved.
We achieve this by recognizing that $n=T_1+T_2+\ldots+T_N$ is a sum of multiple variables $T_j$, corresponding to the partitioning outcome of each electron ($T_j=1$ or 0 if the $j^{\rm th}$ electron is detected at D1 or D2, respectively). 
Instead of $\llangle n^k\rrangle$, we consider the irreducible correlation functions $\llangle T_i T_j \ldots T_k \rrangle$, known as multivariate cumulants in statistics~\cite{Indian1962} or connected diagrams in field theory~\cite{fetter2012quantum}, to quantify the effect of interactions. 
Importantly, if the presence of the $i^{\rm th}$ electron does not influence the $j^{\rm th}$ electron, all multivariate cumulants involving both $T_i$ and $T_j$ will be zero.
Here we focus on the symmetrised multivariate cumulants $\kappa_k$ defined by averaging the cumulants over all possible combinations of exactly $k$ distinct variables $T_j$ out of $N$. 
If electrons are statistically indistinguishable (all placed in the same SAW minimum, or all in different SAW minima), all terms in the averaging are equal and $\kappa_k=\llangle T_1 T_2 \ldots T_k \rrangle$. 
In this case, the multivariate cumulants $\kappa_k$ are entirely determined by the counting probability distribution $P_{(N-n,\,n)}$, and can be computed from both measurements and models (see Methods). 

We now illustrate the meaning of multivariate cumulants using the experimental data from Fig.~\ref{figure2}a-c (where $N=4$) and show the corresponding cumulants ($\kappa_1, \ldots, \kappa_4$) in panels d-f.
The first-order average $\kappa_1=\langle T_1 \rangle = \langle n \rangle/N$ is simply the marginal probability for one electron to be transmitted into D1; it changes monotonously from $0$ to $1$ with detuning parameter $\Delta$.
For a binomial distribution of independent trials, all high-order cumulants $\kappa_{k>1}$ are zero, and this is indeed the case in Fig.~\ref{figure2}d where all 4 electrons are distributed into separate SAW minima. 
To gain intuition about the second-order correlations, we consider the second central moment $\langle n^2 \rangle - \langle n \rangle^2$ of the counting statistics, which is always equal to $N\kappa_1(1-\kappa_1) + N(N-1)\kappa_2$. 
It consists of two terms: the ideal gas contribution proportional to $N$ and representing independent shot-noise accumulation, and the interaction-driven term proportional to $\kappa_2$. 
Coulomb repulsion leads to negative two-body correlations and the corresponding suppression of fluctuations in $n$ (anti-bunching). 
Indeed, we observe $\kappa_2<0$ both in Fig.~\ref{figure2}e for two electron pairs and in Fig.~\ref{figure2}f for a quadruplet. 
The difference between the two cases is revealed by considering higher orders of correlation: while $\kappa_3$ and $\kappa_4$ are close to zero when electrons are restricted to interact in pairs only (panel e), all cumulants $\kappa_k$ up to $k=N$ are generally non-zero when all $N$ electrons are placed in the same SAW minimum (panel f in Fig.~\ref{figure2} for $N=4$ and Extended Data Fig.~\ref{extended-figure2} for $N=3$ to 5). 
Higher-order cumulants oscillate with detuning and exhibit $k-1$ extrema separated by $k-2$ zeros. 
Even cumulants are symmetrical, while odd cumulants are antisymmetrical, reflecting the symmetry of the Y-junction already evident from the partitioning probabilities in panels a-c.

In the following, we show that the observed pattern of high-order correlations aligns with the universal signature of a strongly-correlated liquid, and locate the droplet state within the phase diagram associated with the gas-liquid crossover.

\begin{figure*}[t]
    \includegraphics[scale=0.52]{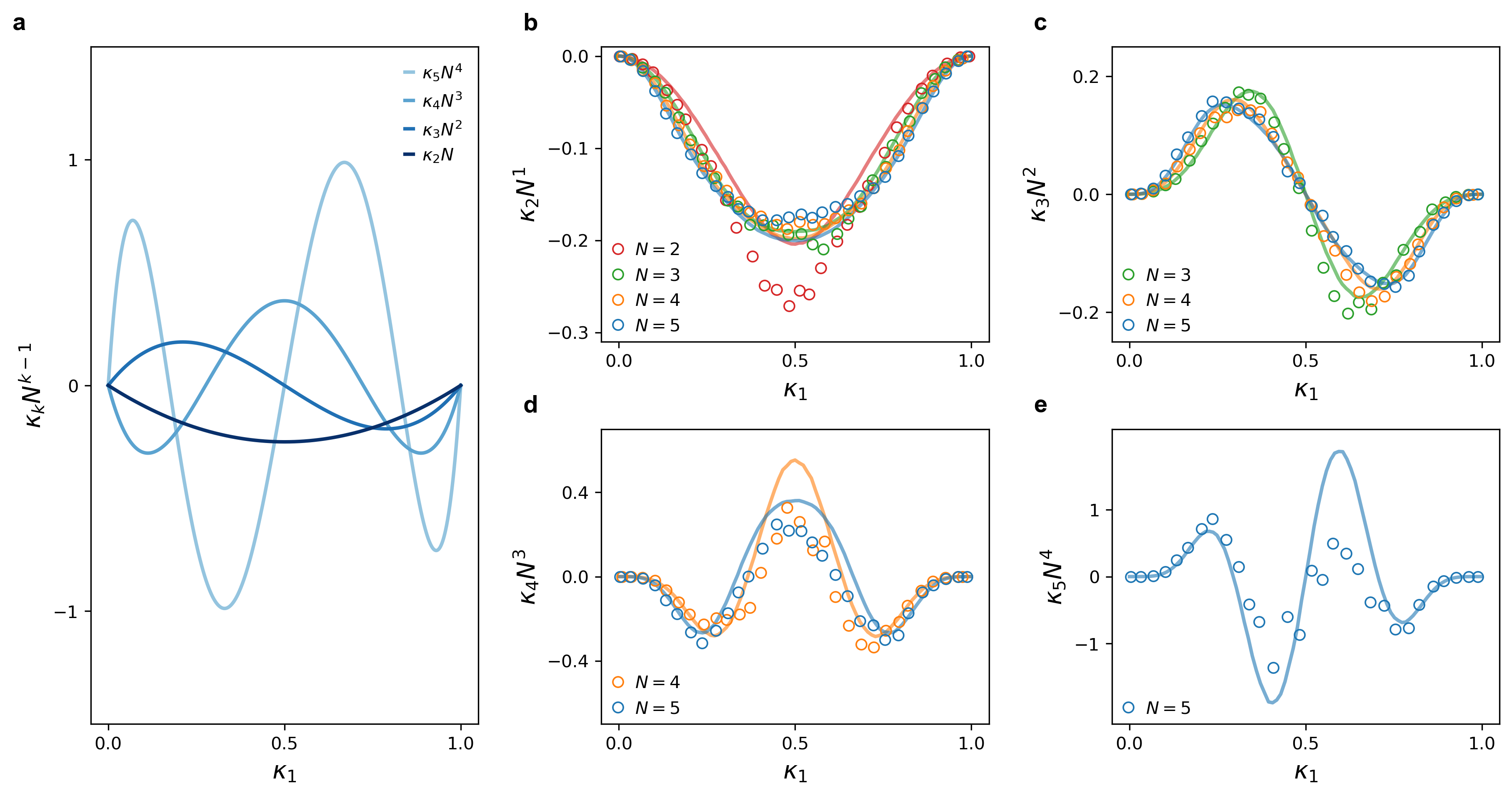}
    \caption{
    \textbf{Scaling of correlation functions with the number of particles.} 
    \textbf{a}, Leading-term universal asymptotics of the repulsion-dominated cumulants in the large $N$ limit, as function of $\kappa_1$ according to Eq.~\eqref{eq:ultraspherical}. 
    \textbf{b-e}, Measured cumulants $\kappa_k$ of order $k=2\ldots 5$ for droplets with $N=k\ldots 5$. Lines show corresponding simulations of sudden partitioning of an equilibrium Coulomb plasma confined in a quartic-parabolic potential with realistic microscopic parameters (see Methods). 
    }
    \label{figure3}
\end{figure*}

\vspace{3mm}
\noindent
\textbf{UNIVERSAL SIGNATURES OF A COULOMB LIQUID}

\noindent
Particle scattering statistics have historically been instrumental in mapping the observed correlations of many-body systems onto phase diagrams of strongly-correlated matter. 
A prime example are relativistic ion collisions used to study the phase diagram of QCD.
In particular, at low baryonic densities~\cite{niida2021signatures}, the transition from quark-gluon plasma at temperatures $T>T_\text{c}$ to hadronic fluid at $T<T_\text{c}$ is not a sharp phase transition but rather a smooth crossover~\cite{aoki2006order,borsanyi2010qcd} with $k_\text{B}T_\text{c} \sim \SI{170}{\mega e\volt}$.
Freeze-out of fluctuations (due to quench of equilibrium during expansion~\cite{harris2024qgp}) determines the cumulants in the number of produced hadrons, which can be used to connect QCD calculations to the statistics from collision experiments~\cite{gupta2011scale}.
In analogy with this high-energy crossover, the relevant state of matter for our electron droplets is a one-component Coulomb plasma~\cite{baus1980statistical} with a crossover from a Coulomb gas at $T>T_{\text{c}}$ to a Coulomb liquid at $T<T_{\text{c}}$.
The crossover temperature $T_{\text{c}}$ is determined by the dimensionless plasma parameter $\Gamma^{(\text{pl})} \sim T_{\text{c}}/T$.
Unlike in QCD, where particle creation occurs, the equilibrium ensemble here is canonical, as no additional particles are generated.

In the ideal gas limit of a Coulomb plasma, $\Gamma^{(\text{pl})} \ll 1$, the multivariate cumulants $\kappa_k$ take universal (albeit trivial) values, $\kappa_{k>1}\to 0$. 
In the opposite limit of an interaction-dominated liquid, $\Gamma^{(\text{pl})} \gg 1$, we can derive a universal form of the large-$N$ scaling of the cumulants $\kappa_k$ from the condition that the distribution of $n$ is governed by Coulomb repulsion rather than by the statistical fluctuations. 
For interactions to dominate the variance of the observable $n$, the interaction term must asymptotically cancel the ideal gas term, hence $\kappa_2 \to -\kappa_1(1-\kappa_1) N^{-1}$. 
Extending the argument by induction to higher $k$, we find that $\kappa_k \propto N^{-k+1}$ with the proportionality coefficient given by a specific parameter-free polynomial of order $k$ in $\kappa_1$, as shown in Fig.~\ref{figure3}a (see Methods). 
As our experiment allows tuning $\kappa_1$, we plot in Fig.~\ref{figure3}b-e the rescaled cumulants $\kappa_k N^{-k+1}$ as functions of $\kappa_1$ for $k$ up to 5, using our experimental data for $N=k\ldots 5$.
We observe that the scaling with $N$ expected in the interaction-dominated limit is obeyed as soon as $N\geq3$.
The pattern and the magnitude of oscillations in panels b-e are also in good qualitative agreement with the universal prediction in panel a, confirming that our droplets are large enough to exhibit the emergent behaviour of a strongly-correlated liquid.

\vspace{3mm}
\noindent
{\textbf{EFFECTIVE ISING MODEL}}

\noindent
For quantitative analysis of our finite-$N$ data in terms of interaction strength and thermodynamic phase diagram, we employ the archetype of classical lattice gas models \cite{pelissetto2000critical} and describe the gas-liquid transition with the Ising model on a complete graph (all-to-all interactions).
This model is defined by the following Hamiltonian, expressed in terms of directly measurable partitioning variables,
\begin{align} \label{eq:Hamiltonian}
    \mathcal{H} = U \sum_{\substack{i,j=1 \\ i \neq j}}^N 
    \left(T_i-\frac{1}{2}\right)\left(T_j-\frac{1}{2}\right) + \mu \sum_{i=1}^{N} T_i 
\end{align}
where $U$ is the interaction strength and $\mu$ controls the upper-lower charge balance within the central channel, with $\mu=0$ corresponding to symmetric partitioning statistics $P_{(N-n,n)}=P_{(n,N-n)}$.
We find that a sudden quench of equilibrium fluctuations governed by the Ising Hamiltonian~\eqref{eq:Hamiltonian} at temperature $T$ accurately reproduces the measured cumulants as function of $\mu=-\alpha(\Delta-\Delta_0)$, as shown by solid lines in Fig.~\ref{figure2}e,f for $N=4$ and in Extended Data Fig.~\ref{extended-figure1} and \ref{extended-figure2} for $N=2$ to 5.
In this model, $U/k_\text{B}T$ and $\Delta_0$ are fitted independently for each $N$, while $\alpha/k_\text{B}T$ is fixed globally ($\alpha$ is the voltage-to-energy conversion factor).
Fitted values are listed in Extended Data Table \ref{extended-table1}.

\begin{figure*}[t]
    \includegraphics[scale=0.62]{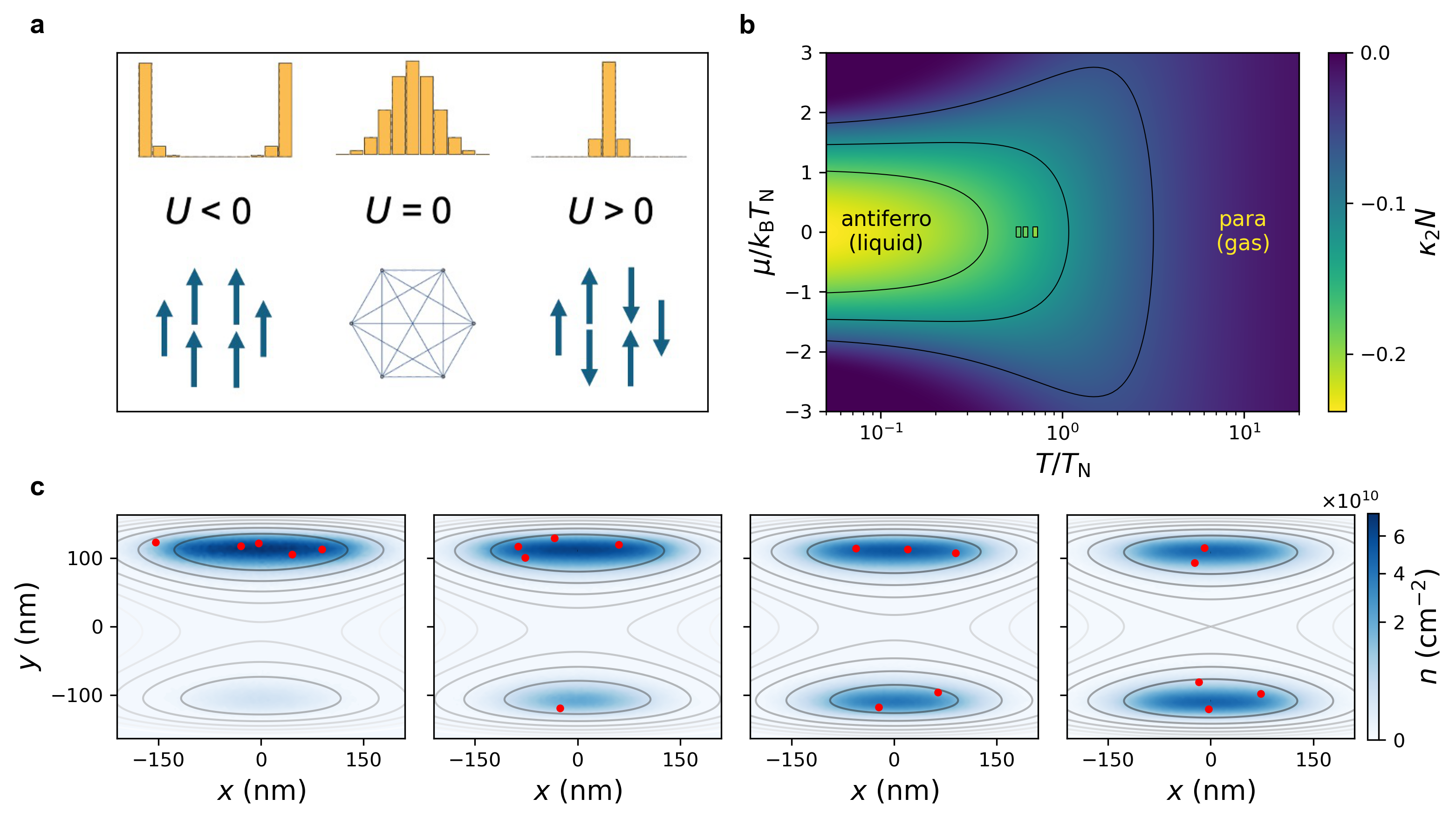}
    \caption{
    \textbf{a}, Interpretation of partitioning in terms of magnetic spin interactions. Uncorrelated partitioning ($U=0$, binomial distribution in the middle), bunching ($U<0$) and anti-bunching ($U>0$) correspond, respectively, to paramagnetic, ferromagnetic and antiferromagnetic phases of the Ising model on a complete graph, for which counting statistics gives the distribution of the total magnetization. 
    \textbf{b}, Phase diagram of the antiferromagnetic crossover in the thermodynamic limit of the Ising model, with appropriately scaled negative pair correlations $\kappa_2 N$ as the order parameter. The axes are given by the temperature $T$ and the magnetic field $\mu$, scaled by the N\'eel temperature $T_{\text{N}}$. The measured correlations for $N=3$, $4$ and $5$ at $\mu=0$ are shown by colors in small squares. The horizontal position $T/T_\text{N}$ of the squares is obtained from the fits of the Ising model to the partitioning curves (Extended Data Table \ref{extended-table1}). The slight deviations in color between the phase diagram and the measured values (36\%, 20\%, and 11\%, respectively) are dominated by the finite-$N$ effect, not by discrepancy with the model.
    \textbf{c}, Four configurations of the 2D confining potential in the central channel (level lines), together with snapshots of the spatial positions of $N=5$ electrons (red dots) from Monte-Carlo simulations of a classical Coulomb plasma (see Methods). The color scale shows the calculated average electron density. The four panels correspond to (from left to right) $\Delta-\Delta_0=101$, 60, 21, and 0~\si{\milli\volt}.}
    \label{figure4}
\end{figure*}

The Ising model establishes a useful analogy between the phases of magnetically interacting spins $s_i=2T_i-1=\pm 1$ and our partitioning statistics, see Fig.~\ref{figure4}a. 
Since the Coulomb plasma is characterised by repulsive interactions ($U>0$), the gas-to-liquid transition corresponds to the paramagnetic-to-antiferromagnetic transition of Ising spins. 
Unlike the usual sharp phase transition in the ferromagnetic case ($U<0$), here it is a crossover in temperature because an all-to-all antiferromagnetic coupling ($U>0$) imposes global correlations but not a particular spin pattern. 
In the large-$N$ limit and in absence of polarizing field ($\mu=0$), the characteristic scale of the antiferromagnetic crossover is the N\'eel temperature $k_\text{B}T_\text{N}=UN/2$ (see Methods) which we identify with the crossover temperature $T_\text{c}$ of the strongly correlated Coulomb liquid. 

The cumulants $\kappa_k$ serve as the irreducible correlation functions of the Ising spins and can be used as order parameters to quantitatively trace the crossover in a phase diagram. 
This is illustrated for $\kappa_2 N$ in Fig.~\ref{figure4}b where the scale $T_\text{N}$ represents the characteristic energy required to destroy the correlated state, by either thermal fluctuations ($T$ axis) or by an external field favoring ferromagnetic order ($\mu$ axis). 
Converting the fitted parameters $U/(k_\text{B} T)$ for $N=3\,\ldots\,5$ to $T/T_{\text{N}}=0.6\,\ldots\,0.7$ allows us to compare the measured correlations to their values in the thermodynamic limit.
In terms of plasma state, we find the results closer to liquid than gas ($T<T_\text{c}$), consistent with the observation of cumulant scaling (see Fig.~\ref{figure3}) characteristic of the liquid state.

Finally, we complement the coarse-grained effective Ising model with an explicit microscopic simulation of the Coulomb plasma~\cite{bedanov1994ordering} under the specific conditions of our experiment. 
Figure~\ref{figure4}c illustrates the case of $N=5$ electrons, interacting through an unscreened Coulomb potential and placed in a quartic-parabolic confining potential for different values of detuning voltage $\Delta$. 
The color represents the canonical probability density drawn from classical Monte-Carlo simulations, and the red dots represent a particular snapshot of the corresponding electron positions.
The simulated partitioning cumulants are shown in Fig.~\ref{figure3}b-e with continuous lines.
As the shape of the confinement potential is computed from an electrostatic modelling (see Methods), the only adjustable parameter is the temperature of the droplet, $T=\SI{25}{\kelvin}$.
We observe that these microscopic simulations are consistent with the observed deviations from the ideal scaling and account for finite-size and temperature effects.
We have verified that the qualitative agreement with the scaling limit (Fig.~\ref{figure3}a) is robust to the choice of the confinement potential and to the competition of Coulomb and exchange correlations, as long as the plasma parameter is large enough.

\vspace{3mm}
\noindent
\textbf{CONCLUSION AND OUTLOOK}

\noindent
Inspired by relativistic ion colliders used to study quark-gluon plasma, we have successfully created a plasma droplet of strongly-correlated electrons on a microchip. 
Confining electrons with a surface acoustic wave enables precise manipulation of the number of interacting particles within the electron-plasma droplet, and a gate-tuneable Y-junction provides deterministic control of the effective impact parameter.

A multivariate cumulant analysis of the partitioning data reveals the formation of a strongly-correlated Coulomb liquid, emerging with as few as three electrons in the droplet.
This mirrors the incompressible liquid of hadrons bound by nuclear forces, but unlike high-energy ion collisions, our approach allows us to trace universal signatures of collectivity at very low particle numbers.

Drawing a powerful analogy with condensed-matter physics, our results align well with an Ising model on a complete graph. 
The observed electron antibunching, governed by Coulomb interactions, can be compellingly interpreted as antiferromagnetic ordering below the N\'eel temperature.

In future, an exciting direction would be to extend this methodology to lower effective temperatures and strong magnetic fields, where quantum Hall states emerge in two-dimensional electron systems~\cite{monarkha2013two}, and have already been simulated for small particle numbers~\cite{lunt2024realization}. 
Notably, evidence of electron bunching in a pair-partitioning experiment~\cite{ubbelohde2014partitioning} under high magnetic fields suggests the potential formation of a Laughlin state droplet~\cite{laughlin1983quantized}, opening new avenues for engineering exotic correlated states in electron systems.

\vspace{3mm}
\noindent
{\bf ACKNOWLEDGEMENTS}

\noindent
The authors acknowledge fruitful discussions with Urs Achim Wiedemann, Davis Zavickis and Benjamin Sac\'ep\'e during the preparation of this manuscript.
This project has received funding from the European Union H2020 research and innovation program under grant agreement No. 862683, “UltraFastNano”. 
C.B., H.S., and J.S. acknowledge funding from the Agence Nationale de la Recherche under the France 2030 programme, reference ANR-22-PETQ-0012. 
C.B. acknowledges financial support from the French Agence Nationale de la Recherche (ANR) project QUABS ANR-21-CE47-0013-01.
J.W. acknowledges the European Union H2020 research and innovation program under the Marie Sklodowska-Curie grant agreement No. 754303. 
M.A. acknowledges the MSCA co-fund QuanG Grant No. 101081458, funded by the European Union and the program QuanTEdu-France n° ANR-22-CMAS-0001 France 2030.
L.M. acknowledges the program QuanTEdu-France n° ANR-22-CMAS-0001 France 2030.
T.V. acknowledges funding from the French Laboratory of Excellence project ``LANEF'' (ANR-10-LABX-0051).
A.D.W. and A.L. thank the DFG via ML4Q EXC 2004/1 - 390534769, the BMBF-QR.X Project 16KISQ009 and the DFH/UFA Project CDFA-05-06.
E.P., R.S. and V.K. have been supported by Latvian Quantum Initiative within European Union Recovery and Resilience Facility project no.\ 2.3.1.1.i.0/1/22/I/CFLA/001 and grant no.\ lzp2021/1-0232 from the Latvian Council of Science.
 
Views and opinions expressed are those of the authors only and do not necessarily reﬂect those of the European Union or the granting authority. Neither the European Union nor the granting authority can be held responsible for them.

\vspace{3mm}
\noindent
{\bf AUTHOR CONTRIBUTIONS}

\noindent
J.S. performed the experiment with expert help from J.W. and technical support from T.V., M.A., L.M, C.G and M.U. \, J.W. fabricated the sample. E.P. performed the cumulant calculations and R.S. the Monte Carlo simulations with guidance from V.K.\, A.L and A.D.W. provided the high-quality GaAs/GaAlAs heterostructure. J.S, E.P, C.B, V.K. and H.S. wrote the manuscript with feedback from all authors. H.S and C.B. supervised the experimental work. V.K. conceptualised the theoretical interpretation. C.B. has initiated the project.

\newpage
\noindent
{\bf METHODS}

\vspace{3mm}
\noindent
{\bf Device description}

\noindent
The device is fabricated in a Si-doped GaAs/AlGaAs heterostructure grown by molecular beam epitaxy. 
The two-dimensional electron gas (2DEG) resides \SI{110}{\nano\meter} below the surface, with electron density  
\SI{2.8e11}{\per\square\centi\meter} and mobility \SI{9e5}{\centi\meter\squared\per\volt\per\second}. Metallic gates (Ti, \SI{3}{\nano\meter}; Au, \SI{14}{\nano\meter}) are deposited on the surface of the semiconductor using electron-beam lithography. 
All measurements are performed at a temperature of about \SI{20}{\milli\kelvin} in a $^3$He/$^4$He dilution refrigerator. 
The sample and measurement scheme are the same as in Ref.~\cite{wang2023coulomb}. 
A set of negative gate voltages is applied to the surface gates to deplete the 2DEG underneath and create the nanostructures, including 4 quantum dots (QD), 4 quantum point contacts (QPC), and 2 guiding rails which are fully depleted. 
These rails connect the source QDs to the detector QDs and merge in the centre to form a single \SI{40}{\micro\meter}-long channel, equipped with a narrow barrier gate in the middle to tune the shape of the confining potential from a single well to a double well.

\vspace{3mm}
\noindent
{\bf SAW generation}

\noindent
The surface acoustic wave (SAW) is generated using a double-finger interdigital transducer (IDT) deposited on the surface and placed at a distance of \SI{1.5}{\milli\meter} from the device.
The metallic fingers are fabricated using electron-beam lithography and thin-film evaporation (Ti, \SI{3}{\nano\meter}; Al, \SI{27}{\nano\meter}) on the heterostructure.
The IDT consists of 111 cells with a periodicity of 1\,µm and a resonance frequency \SI{2.86}{\giga\hertz} at low temperature.
The aperture of the transducer is \SI{50}{\micro\meter}.
To perform electron transport by SAW, a radiofrequency signal is applied on the IDT at its resonance frequency for a duration of \SI{60}{\nano\second}.
To have a strong SAW confinement potential, the signal is amplified to \SI{28}{\decibel} using a high-power amplifier before being injected into a coaxial line of the cryostat through a series of attenuators.
The velocity of the SAW is \SI{2860}{\meter\per\second}.

\vspace{3mm}
\noindent
{\bf Electron transfer}

\noindent
Each single-shot experiment corresponds to the transfer of one or a few electrons from the source QDs to the detector QDs using the SAW as transport carrier.
To prepare a given number $N$ of electrons in a source QD, we employ a sequence of fast voltage pulses to the channel gate and reservoir gate controlling the tunnel barriers of the QD.
This sequence consists of three steps: initializing the QD, loading the electrons into the QD, and preparing the QD for electron transfer.
To initialise the source QD, electrons previously present in the QD are removed. 
Then, a given number of electrons is loaded into the QD by accessing a particular loading position in the charge stability diagram of the QD. 
Finally, these electrons are trapped within the QD by switching to a holding configuration, from where they will be taken away by the SAW. 
At the same time, the two detector QDs are set in a configuration for which the electrons transported by the SAW will be captured with high fidelity. 
For more details, see Supplementary Note~\ref{supp:transfer}.
By sensing the QPC currents of both the source and detector QDs with single-electron resolution, and comparing their values before and after the electrons are transferred by the SAW, the precise number of electrons transferred to each detector can be determined.  
When calculating the partitioning probability, the very few events that do not conserve the total number of electrons are excluded by a post-selection routine.

In our experiment, the electrons are deterministically loaded into specific locations within the SAW train. 
The plunger gate of the QD is used to trigger the sending of the electrons into a precise minimum of the periodic SAW potential with a \SI{30}{\pico\second} resolution.
This precise control allows for the formation of an electron droplet containing up to 5 electrons, using the two source QDs of the device.
To synchronise the two trigger pulses with the radio-frequency  signal generating the SAW, we use two arbitrary waveform generators (AWG) combined with a synchronization module. 
The outputs of the AWGs are connected to the plunger gates via high-bandwidth bias tees for voltage pulsing and dc-biasing.

\vspace{3mm}
\noindent
{\bf Electron partitioning}

\noindent
The electron droplet is partitioned at the Y-junction located at the end of the central channel, after a flight time of \SI{14}{\nano\second}.
By applying a voltage detuning $\Delta=V_\text{U}-V_\text{L}$, where $V_\text{U}$ and $V_\text{L}$ are the voltages applied to the side gates 
of the central channel, we can control the partitioning ratio between the two detectors D1 and D2.  
For all partitioning experiments reported here, the barrier gate voltage is set to $V_\text{B}=\SI{-1.25}{\volt}$ in order to have a single central channel with a weak double-well potential profile. Careful analysis of the double-well potential and the electron number equilibration in the central channels is described in Supplementary Notes \ref{supp:tuning} and \ref{supp:indistinguishability}.

\vspace{3mm}
\noindent
{\bf Symmetrised multivariate cumulants}

\noindent
In our experiment, the observable is the number $n$ of electrons measured at the detector D1, which can be expressed as a sum of binary variables $T_j$.
From $n=\sum_{j=1}^{N} T_j$ and $T_j^2=T_j$, we derive the general relation
\begin{align}
    m_k =\binom{N}{k}^{-1} \sum_{n=k}^N \binom{n}{k}  p_n \label{eq:mpviap}
\end{align}
between the probabilities $p_n=P_{(N-n,n)}$ of the full counting statistics (FCS) \cite{levitov1996electron} and the $k$-th order symmetrised multivariate moments $m_k$ defined as averages of all permutations of $k$ distinct variables,
\begin{align} 
    m_k = \binom{N}{k}^{-1} \sum_{
    \substack{ 1 \leq j_1 < j_2  <  \ldots < j_k \leq N}} 
    \langle T_{j_1} T_{j_2} \ldots T_{j_k} \rangle \, ,
    \label{eq:mpdef}
\end{align}
where $\binom{N}{k}=N!/[k! (N-k)!]$ is the binomial coefficient.
The corresponding symmetrised multivariate cumulants
\begin{align}
    \kappa_k=
    \binom{N}{k}^{-1} \sum_{
    \substack{ 1 \leq j_1 < j_2  <  \ldots < j_k \leq N}} 
   \llangle T_{j_1} T_{j_2} \ldots T_{j_k}  \rrangle
    \label{eq:kappadf}
\end{align}
are, in general, not uniquely determined by FCS probabilities, and their calculation requires additional information (such as symmetry constraints or a microscopic model). 

For statistically equivalent particles (i.e.\ full permutational symmetry of the multivariate probability distribution), all terms in Eq.~\eqref{eq:mpdef} and in Eq.~\eqref{eq:kappadf} are equal, and the moments $m_k$ can be related to cumulants $\kappa_k$ via standard univariate relations~\cite{mccullagh2018tensor}, $\ln (1+\sum_{k=1}^{\infty} m_k \, z^k/k!)=\sum_{k=1}^{\infty} \kappa_k \, z^k/k!$. Using an explicit formula in terms of Bell polynomials~\cite{comtetbook}, we can write
\begin{align}
    \kappa_k = \sum_{j=1}^k (j-1)! (-1)^{j-1} B_{kj}
    \left( m_1 , m_2 , \ldots , m_{k-j+1} \right) \, .
    \label{eq:kappaGeneral}
\end{align}  
See Supplementary Note~\ref{supp:cumulantsGeneral} for derivation of Eq.~\eqref{eq:mpviap} and explicit formulas for $\kappa_k$ for $k=1$ to $5$. 
An example of correlated partitioning, where Eq.~\eqref{eq:kappaGeneral} is not valid and the general combinatorial expressions for multivariate cumulants~\cite{Gardiner1986HandbookOS,Indian1962} need to be used, is illustrated in Fig.~\ref{figure2}e and described in detail in Supplementary Note~\ref{supp:2e2epartitioning}. 

There is an important distinction between our method for extracting interaction signatures and the approach of so-called factorial cumulants considered in the context of electron transport~\cite{kambly2011factorial} and particle physics~\cite{Kitazawa2017}. 
The multivariate moments defined by Eq.~\eqref{eq:mpdef} can be written $m_k = \langle (n)_k \rangle/(N)_k$ where $(x)_k=x (x-1) \times \ldots \times (x-k+1)$ is the falling factorial and $\langle (n)_k \rangle$ is known as the factorial moment of FCS~\cite{kambly2011factorial}. 
The $k$-dependent denominator $(N)_k$ in this expression for $m_k$ makes the $\kappa_k$ distinct from the factorial cumulants, see additional discussion in Supplementary Note~\ref{supp:cumulantsGeneral}.

\vspace{3mm}
\noindent
{\bf Ising model on a complete graph}

\noindent
The Ising model on a complete graph is exactly solvable~\cite{Baxter2016} and hence equilibrium fluctuations at any freeze-out quench temperature $T$ can be computed for any $N$.
The Ising Hamiltonian of Eq.~\eqref{eq:Hamiltonian} can be expressed as a quadratic form of the observable $n=\sum_{j=1}^{N} T_j$, 
\begin{align} \label{eq:Hamiltonian2}
    \mathcal{H} = U n^2 + (\mu-NU) n + U N(N-1)/4 \, .
\end{align}
The corresponding exact counting statistics in a canonical ensemble is $p_n=c_n/Z$ with the partition function $Z=\sum_{n=0}^{N} c_n$ and the statistical weights
\begin{align} \label{eq:cn}
    c_n = \binom{N}{n} e^{-\beta U n (n-N) - \beta \mu n} \, ,
\end{align}
where $\beta=1/k_\text{B}T$. 
Together with Eq.~\eqref{eq:mpviap} and Eq.~\eqref{eq:kappaGeneral}, this gives a way to calculate the exact multivariate cumulants $\kappa_k$ of all orders $k \leq N$ at any $N$.

In order to make the connection with the thermodynamic phase diagram in terms of $\mu$ and $T$ in the large-$N$ limit, explicit analytic expressions are obtained following Ref.~\cite{Baxter2016}.
We apply the lowest-order Stirling's formula $m!\approx m^{m} e^{-m} \sqrt{2 \pi m}$ to the factorials in the binomial coefficient $\binom{N}{n}$ in Eq.~\eqref{eq:cn} and perform expansion of $\ln(c_n)$ near its maximum $n \approx \langle n \rangle$ up to quadratic order. 
This results in a Gaussian approximation to $p_n$ of the form
\begin{align} \label{eq:Gaussian}
    p_n \propto e^{ -(\beta +\beta') \, U \, (n - \kappa_1 N)^2 } \, ,
\end{align}
where $\beta' = \left[ 4\kappa_1(1-\kappa_1) k_\text{B} T_\text{N} \right]^{-1}$ and $k_\text{B} T_\text{N} = U N/2$ is the zero-field N\'eel temperature for the antiferromagnetic crossover.
The relation between the effective magnetic field $\mu$ and the effective magnetization $\kappa_1=\langle n \rangle /N$ in the large-$N$ limit is given by the transcendental equation~\cite{bragg1934effect} 
\begin{align} \label{eq:Bragg}
    2\kappa_1 - 1 = \tanh{\left[-\frac{T_{\text{N}}}{T}\left(2\kappa_1 - 1 - \frac{1}{2}\frac{\mu}{k_{\text{B}}T_{\text{N}}}\right)\right]}  \, ,
\end{align}
which has only one solution for the antiferromagnetic sign of the coupling ($U>0$).

To quantify the antiferromagnetic correlations in the thermodynamic limit, we choose the pair correlation function $\llangle T_1 T_2 \rrangle =\kappa_2$ as the order parameter. 
It is obtained from the identity $\langle n^2 \rangle - \langle n \rangle^2 = N\kappa_1(1-\kappa_1) + N(N-1)\kappa_2$ in the large-$N$ limit. 
Treating $n$ as a continuous variable and using the Gaussian approximation of Eq.~\eqref{eq:Gaussian}, this gives the leading-order behaviour of $\kappa_2$ at a fixed $T/T_{\text{N}}$ and $N\to\infty$,
\begin{align} \label{eq:Isingscaling}
    \kappa_2 N = - \, \frac{4 \kappa_1^2 (1-\kappa_1)^2}{4 \kappa_1 (1-\kappa_1)+T/T_{\text{N}}} \, .
\end{align}
A numerical solution to Eq.~\eqref{eq:Bragg} together with Eq.~\eqref{eq:Isingscaling} is used for the phase diagram in Fig.~\ref{figure4}b.

As there is no lattice on a full graph favoring a particular pattern of staggered magnetization, the antiferromagnetic transition here is not a second-order phase transition but a crossover. 
The corresponding change in free energy has a weaker divergence ($\log N$) in the thermodynamic limit than at the ferromagnetic transition.
The corresponding singular part \cite{pelissetto2000critical} of the free energy change between $T=\infty$ and $T \to 0^{+}$ is $\beta\Delta F_{U>0} = (1/2)\ln(1+T_{\text{N}}/T)$.

On the ferromagnetic side ($U<0$), we note that $\kappa_2\,N$ diverges when the temperature $T$ approaches the Curie temperature $T_\text{C}=-T_{\text{N}}>0$ as $\kappa_2 N \propto (T-T_\text{C})^{-1}$ for $T\to T_\text{C}^{+}$. 
At $T \leq T_\text{C}$, the Gaussian approximation of Eq.~\eqref{eq:Gaussian} breaks down and strong ferromagnetic order sets in. 
This corresponds to the droplet scattering at the Y-junction as a whole (without partitioning), with probability $\kappa_1$ to go to detector D1 and with $\kappa_2 =\kappa_1(1-\kappa_1)>0$ in the large-$N$ limit. 
For a large but finite droplet, there is no symmetry breaking, hence $\kappa_{k>1} = O(1)$ unlike $O(N^{-k+1})$ in the antiferromagnetic case.

\vspace{3mm}
\noindent
{\bf Universal scaling of partitioning cumulants}

\noindent
The interaction-dominated partitioning of a large droplet at $U>0$ is described by the antiferromagnetic phase of the effective Ising model with $T/T_{\text{N}} \to 0$ and $N \to \infty$. 
The Boltzmann factor in Eq.~\eqref{eq:cn} suppresses the fluctuations of $n$ around $\langle n \rangle = \kappa_1 N$ and caps the large-$N$ asymptotics of univariate cumulants from $\llangle n^k \rrangle = O(N)$ (Gaussian limit of binomial distribution) to $\llangle n^k \rrangle = O(1)$. 
From the latter condition (which is independent of the specifics of the Ising model), we derive the asymptotics $\kappa_k = G_k(\kappa_1) N^{-k+1} + O(N^{-k})$ for $k \ge 2$ where the prefactor
\begin{align} \label{eq:ultraspherical}
    G_k(\kappa_1) = -\frac{(k-1)!}{2} C^{(-1/2)}_k(2 \, \kappa_1-1) 
\end{align}
is universal and given by the ultraspherical (Gegenbauer) polynomials $C^{(a)}_k$ of degree $k$ and parameter $a=-1/2$. 
The first polynomials up to $k=5$ are plotted in Fig.~\ref{figure3}a, to show the universal strong-correlation asymptotics of the scaled cumulants $\kappa_k N^{k-1}$. 
Note that $G_2=-\kappa_1(1-\kappa_1)$ is also the zero-temperature limit of Eq.~\eqref{eq:Isingscaling}.

The polynomials $G_k(\kappa_1)$ have exactly $k-2$ zeros for $0 < \kappa_1 < 1$, which explains the observed oscillation pattern and provides an exact specific example of  oscillations in high-order cumulants~\cite{flindt2009universal}. 
We note that a similar generic $N^{-k+1}$ scaling has been discussed for cumulants of initial density perturbations in heavy-ion collisions~\cite{wiedemann2014} where it arises for different reasons (dominance of auto-correlations in independent point sources model).

In contrast to antiferromagnetic correlations decaying with $N$ as $\kappa_k \propto N^{-k+1}$, the fluctuations in the ferromagnetic case  are between $n=0$ and $n=N$ only, hence  $\kappa_k = O(1)$, and the limiting form for the unpartitioned scattering ($T/T_\text{C}\to 0$ in the Ising model) is the polynomial $\kappa_k = -\text{Li}_{1-k}\left(\frac{\kappa_1}{\kappa_1-1}\right)$ where $\text{Li}$ is the polylogarithm.

\vspace{3mm}
\noindent
{\bf Coulomb liquid simulations}

\noindent
We model a finite droplet of Coulomb plasma in 2D using a confining single-electron potential $V_{1e}$ and an unscreened Coulomb potential~\cite{bedanov1994ordering} which results in the total potential
\begin{align} \label{eq:UinteractionDef}
    U(\bm{r}_1,\ldots,\bm{r}_N) = \sum_{i=1}^N V_{1e}(\bm{r}_i) 
    + \sum_{i<j} \frac{e^2}{4\pi\epsilon_0\epsilon_r |\bm{r}_i-\bm{r}_j|} \, ,
\end{align}
where $\bm{r}_i=(x_i,y_i)$ is the in-plane coordinate of the $i$-th electron and $\epsilon_r=12.1$ is the relative dielectric permittivity in GaAs. 
The equilibrium distribution of electron coordinates is determined by a classical canonical ensemble at an effective temperature $T$.
We sample electron positions $\{\bm{r}_i\}$ using a random walk Metropolis Monte-Carlo algorithm designed to sample the canonical distribution.
The convergence of corresponding Markov chain is controlled by the Kolmogorov-Smirnov test~\cite{robert1999monte}. 
For each set of parameters, a statistics of positions is collected with the estimated effective sample size ranging from $10^3$ to $10^5$ depending on parameters.
The statistics of positions is translated to partitioning statistics of a sudden quench using binary variables $T_i=\Theta(y_i)$, where $\Theta$ is the Heaviside step function. 
This corresponds to an observable $n=T_1+\ldots+T_N$  counting the number of particles in the $y>0$ half-plane.

The confining electrostatic potential of our experiment can be approximated as a double-well quartic-parabolic 2D potential
\begin{align}
    V_{1e}(\bm{r}) = V_b + \mu_q \frac{y}{y_0} - 8V_b \frac{y^2}{y_0^2} + 16V_b \frac{y^4}{y_0^4} + \frac{m \omega_x^2 x^2}{2} \, , 
    \label{eq:V1eDef}
\end{align}
where $V_b$ is the height of the central barrier, $y_0$ is the distance between the two minima, $\mu_q$ is the transverse energy detuning proportional to the side-gates voltage difference $\Delta-\Delta_0$ (which controls the partitioning of the droplet), and $\omega_x$ is the oscillation frequency in the longitudinal direction, resulting from the confinement potential of the SAW.

The barrier height $V_b=\SI{27.5}{\milli\electronvolt}$ and the distance between minima $y_0=\SI{220}{\nano\meter}$ are estimated from an electrostatic simulation of the gate-controlled potential as explained in Supplementary Note~\ref{supp:tuning}.
The transverse oscillation frequency in the two potential minima is then calculated as $\omega_y=\sqrt{32V_b/(my_0^2)}=\SI{7.0}{THz}$ using the effective mass $m=0.067m_e$ for electrons in GaAs.
The longitudinal oscillation frequency in the SAW potential is estimated from the peak-to-peak amplitude $A_{\text{SAW}}=\SI{42}{\milli\electronvolt}$ and the wavelength $\lambda_{\text{SAW}}=\SI{1}{\micro\meter}$, using the relation $\omega_x = (\pi/\lambda_{\text{SAW}})(2A_{\text{SAW}}/m)^{1/2} = \SI{1.5}{THz}$ (see Supplementary Note 4 in Ref.~\cite{wang2023coulomb}). 
The aspect ratio of the 2D confinement is thus $\omega_x/\omega_y = 0.21$. 

The potential being entirely determined, the effective electron temperature $T$ is the only free parameter to be chosen for good agreement with the experimental data, as shown by solid lines in Fig.~\ref{figure3}b-e using $T=\SI{25}{\kelvin}$.
This value is also consistent with the one extracted from the barrier-height dependence of the thermally-activated hopping rate between the two wells of the quartic potential, as estimated in  Supplementary Note~\ref{supp:tuning}.

\newpage
\noindent
{\bf EXTENDED DATA FIGURES AND TABLES}

\renewcommand{\figurename}{Extended Data Fig.}
\stepcounter{masterfig}
\renewcommand{\tablename}{Extended Data Table}
\renewcommand{\thetable}{\arabic{table}}
\stepcounter{mastertab}

\begin{figure*}[h]
    \includegraphics[scale=0.92]{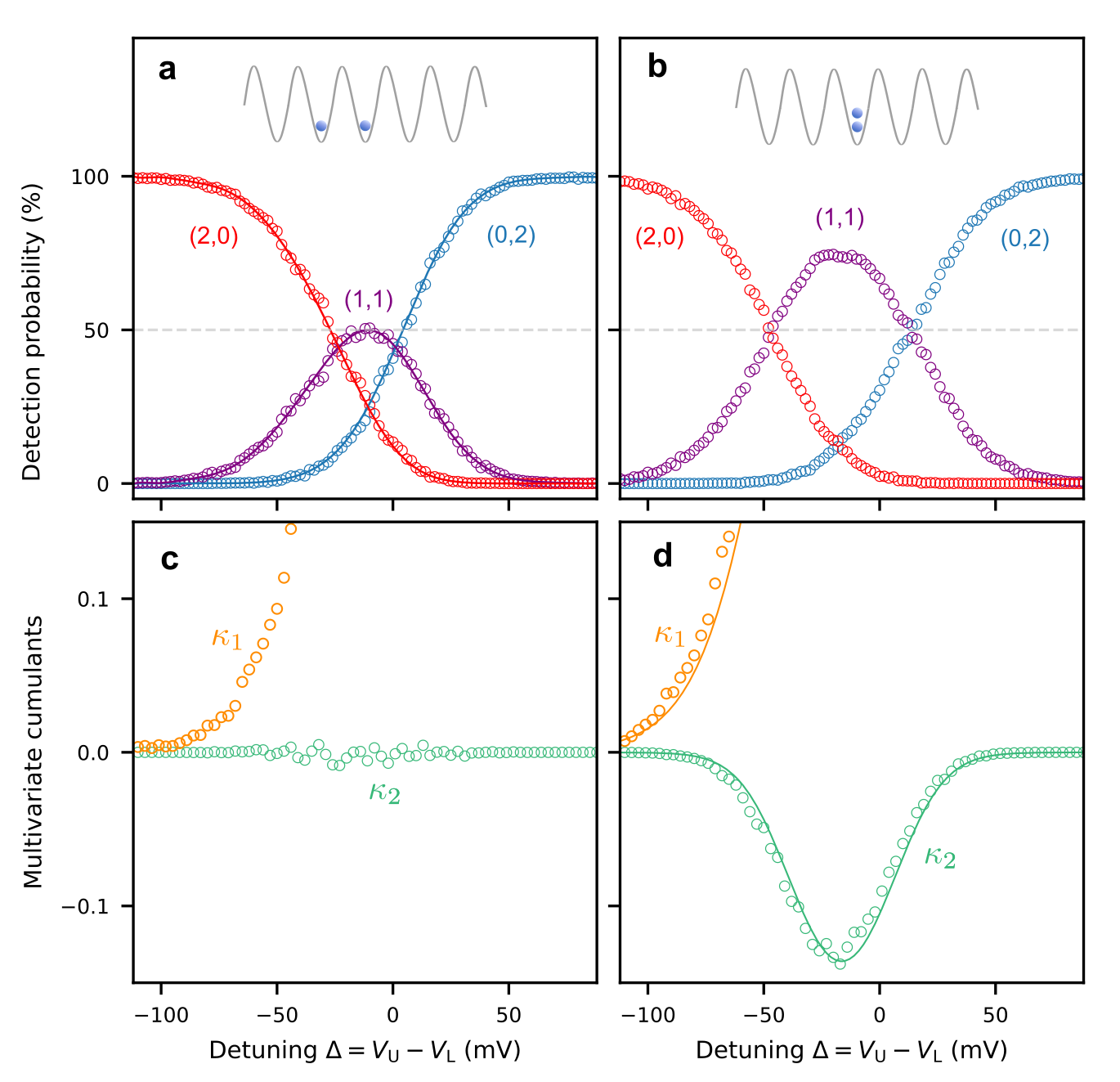}
    \caption{\textbf{Experimental data for partitioning of $N=2$ electrons.} 
    \textbf{a}, Partitioning probabilities when the two electrons are distributed in two different minima and are uncorrelated. 
    \textbf{b}, Partitioning probabilities when both electrons are in the same SAW minimum and are interacting. 
    Panels \textbf{c} and \textbf{d} display the multivariate cumulants corresponding to \textbf{a} and \textbf{b}, respectively. 
    Lines in \textbf{a} are reconstructions using single-electron partitioning data, and lines in \textbf{d} are fitting curves from the Ising model using the parameters given in Extended Data Table~\ref{extended-table1}.}
    \label{extended-figure1}
\end{figure*}

\begin{figure*}[h]
    \includegraphics[scale=0.92]{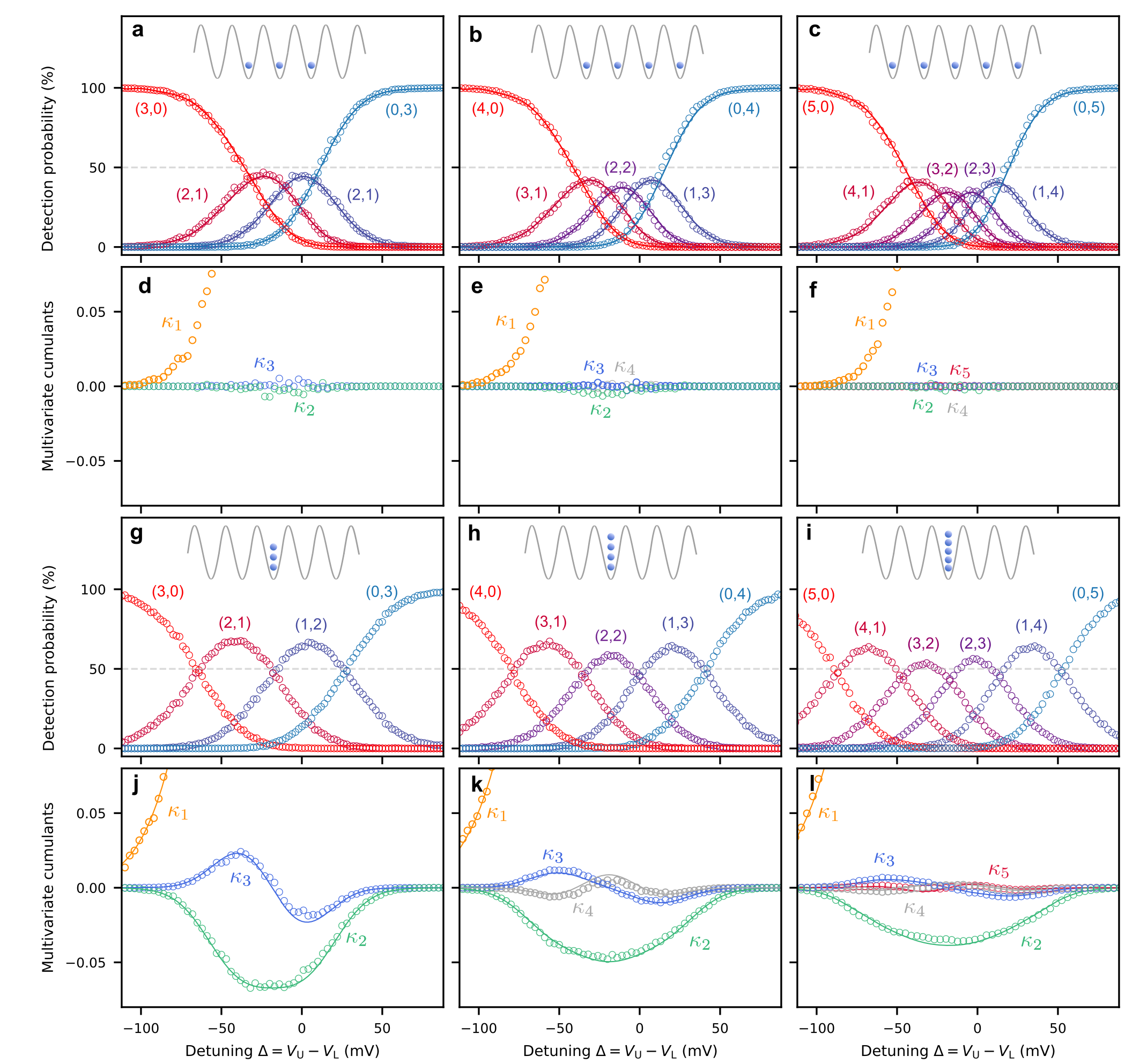}
    \caption{\textbf{Experimental data for partitioning of $N=3$, $4$, $5$ electrons.}
    Panels \textbf{a} to \textbf{f} show the partitioning probabilities and the corresponding multivariate cumulants for electrons distributed across different SAW minima (uncorrelated electrons). 
    Panels \textbf{g} to \textbf{l} show the same quantities when all electrons are placed in the same SAW minimum (interacting electrons). 
    Lines in \textbf{a-c} are reconstructions using single-electron partitioning data, and lines in \textbf{j-l} are fitting curves from the Ising model using the parameters given in Extended Data Table~\ref{extended-table1}.}
    \label{extended-figure2}
\end{figure*}

\newpage
\begin{table}[h]
    \centering
    \begin{tabular}{c|c|c|c|c}
        $N$ & $\Delta_0$ (mV) & $\alpha/k_\text{B}T$ (mV$^{-1})$ & $U/k_\text{B}T$ & $T/T_{\text{N}}$ \\
        \hline
        1  & -12.5  & 0.064 & --   & --   \\
        2  & -16.4  & 0.064 & 1.22 & 0.82 \\
        3  & -18.5  & 0.064 & 0.94 & 0.71 \\
        4  & -19.0  & 0.064 & 0.80 & 0.63 \\
        5  & -17.8  & 0.064 & 0.70 & 0.57 \\
    \end{tabular}
    \caption{\textbf{Fitting parameters of the Ising model for partitioning statistics}. The parameters $\Delta_0$, $\alpha /k_\text{B}T$ and $U/k_\text{B}T$ correspond to the partitioning experiments. The last column shows corresponding values of $T/T_{\text{N}} = 2k_\text{B}T/UN$.}
    \label{extended-table1}
\end{table}

\clearpage
\noindent
{\bf REFERENCES}
\bibliography{references}

\stepcounter{masterfig}
\stepcounter{mastertab}
\stepcounter{mastereq}
\clearpage
\include{Supplementary}

\end{document}

%% file: Supplementary.tex
\renewcommand{\figurename}{Fig.}
\renewcommand{\thefigure}{S\arabic{figure}}
\renewcommand{\tablename}{Table}
\renewcommand{\thetable}{S\arabic{table}}
\renewcommand{\theequation}{S.\arabic{equation}}

\renewcommand{\thesection}{\arabic{section}}
\renewcommand{\thesubsection}{\arabic{section}.\arabic{subsection}}
\renewcommand{\thesubsubsection}{\arabic{section}.\arabic{subsection}.\arabic{subsubsection}}

\makeatletter
\renewcommand{\p@subsection}{}
\renewcommand{\p@subsubsection}{}
\makeatother

\newcommand{\fcircle}{\tikz\fill[black] (0,0) circle (.4ex);}
\newcommand{\hcircle}{\tikz\draw (0,0) circle (0.37ex);}

\setlength{\abovedisplayskip}{8pt} 
\setlength{\belowdisplayskip}{8pt} 

\def\einr{2mm}
\def\spazi{-1.7mm}

\vfill
\begin{center}
    \textbf{\large 
    SUPPLEMENTARY MATERIAL \\
    \vspace{5mm}
    for \\
    \vspace{5mm}
    Evidence of Coulomb liquid phase in few-electron droplets
    }
\end{center}
\vfill
\tableofcontents
\thispagestyle{empty}
\vfill

\pagebreak
\section{CONTROL OF MULTI-ELECTRON TRANSFER}
\label{supp:transfer}

To prepare a droplet of several electrons, precise control over electron placement in a selected potential minimum of the surface acoustic wave (SAW) is essential. 
This task is not straightforward and several parameters must be meticulously tuned and optimised. 
The procedure is described here.

\subsection{Preparation and sending procedure}

\begin{figure}[h]
    \centering
    \includegraphics[scale=0.9]{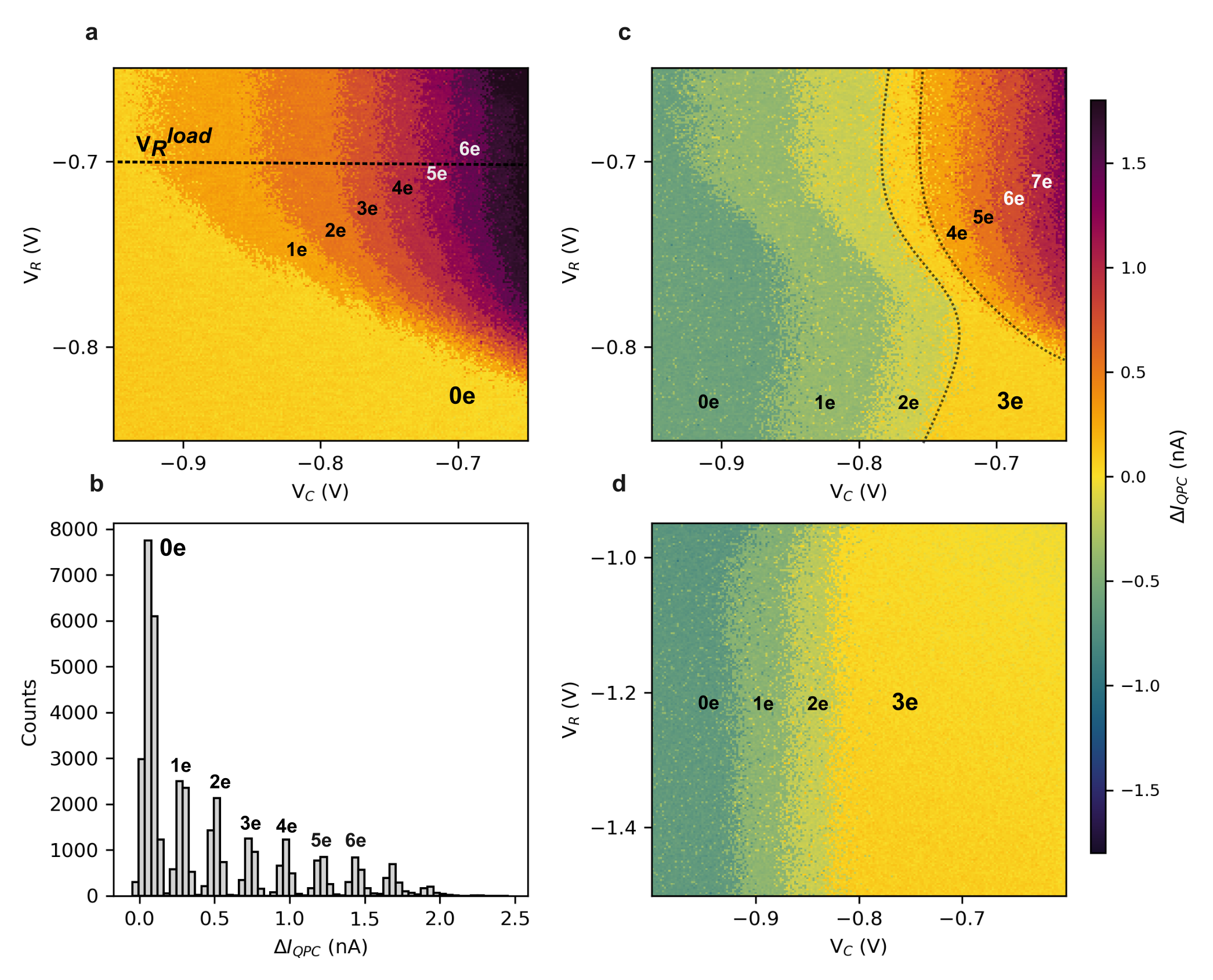}
    \caption{\textbf{Loading and holding maps of a source quantum dot.} 
    \textbf{a}, Loading map of the electron source S1. The axis $V_\text{R}$ and $V_\text{C}$ denote the reservoir (R) and channel (C) gate voltages of the source QD. Each point on the map shows the change in the current through the charge sensor (QPC) after sweeping the gates from a reference position to the gate values of that point and returning to the reference position. Each colour corresponds to a different number of electrons in the QD. The initial electron number before the sweep is zero ($N^\text{init}=0$). The dotted line at $V_\text{R}^\text{load}=\SI{-0.70}{\volt}$ marks the set-point value for electron loading. 
    \textbf{b}, Histogram of QPC currents from the map in \textbf{a}, showing well-separated peaks. 
    \textbf{c}, Same map as in \textbf{a} but for an initial electron number $N^\text{init}=3$. Dotted lines delimit the region (yellow) where the 3 electrons remain trapped in the dot. This map can be viewed as a holding map.
    \textbf{d}, Holding map for $N^\text{init}=3$ in the gate range of the sending configuration.
    }
    \label{figure_S1}
\end{figure}

First, the two source quantum dots (QD) are initialised with the desired number of electrons, while the two detector QDs are left empty. 
The number of electrons in each QD is identified with the help of their respective loading map as shown in Fig.~\ref{figure_S1}a. 
This map is obtained by measuring the current change through the quantum point contact (QPC) coupled to the QD, in a reference configuration of the gate voltages.
This charge sensing method provides a very accurate count of the QD electrons, as illustrated in Fig.~\ref{figure_S1}b. 
Each pixel in the loading map represents the change in QPC current after a specific loading process controlled by the voltages $V_\text{R}$ and $V_\text{C}$ applied to the reservoir gate (R) and channel gate (C) of the source QD \cite{takada2019sound}. 

Then, the system is configured to a holding state, in which the loaded electrons remain trapped in the QDs until the SAW arrives. 
The yellow regions of the holding maps in Fig.~\ref{figure_S1}c,d represent the various sets of gate voltages for which the three electrons initially loaded in the source QD remain trapped.
The holding configuration for SAW transport is when $V_\text{R}$ is around \SI{-1.2}{\volt}.

About one microsecond after the SAW is launched at the IDT, the SAW train reaches the source QDs and the prepared electrons are transported across the device by the piezoelectric potential of the wave.
Using ultra-short RF pulses, we can precisely control the electron distribution within the SAW train, with either all electrons in the same potential minimum, or just a subset in the same minimum, or all in different minima. 
The two main configurations are detailed below.

\subsection{Placing all electrons in the same SAW minimum}

We discuss here the procedure for transporting all the electrons together in the same SAW minimum, thereby forming a single electron droplet. 
The sending process is configured independently for each source using a single detector (D1 for S1 and D2 for S2). 
For this purpose, a large detuning voltage $\Delta$ is applied to the side gates to tilt the confinement potential of the central channel and direct all the electrons towards the selected detector. 

\begin{figure}[h]
    \centering
    \includegraphics[scale=0.9]{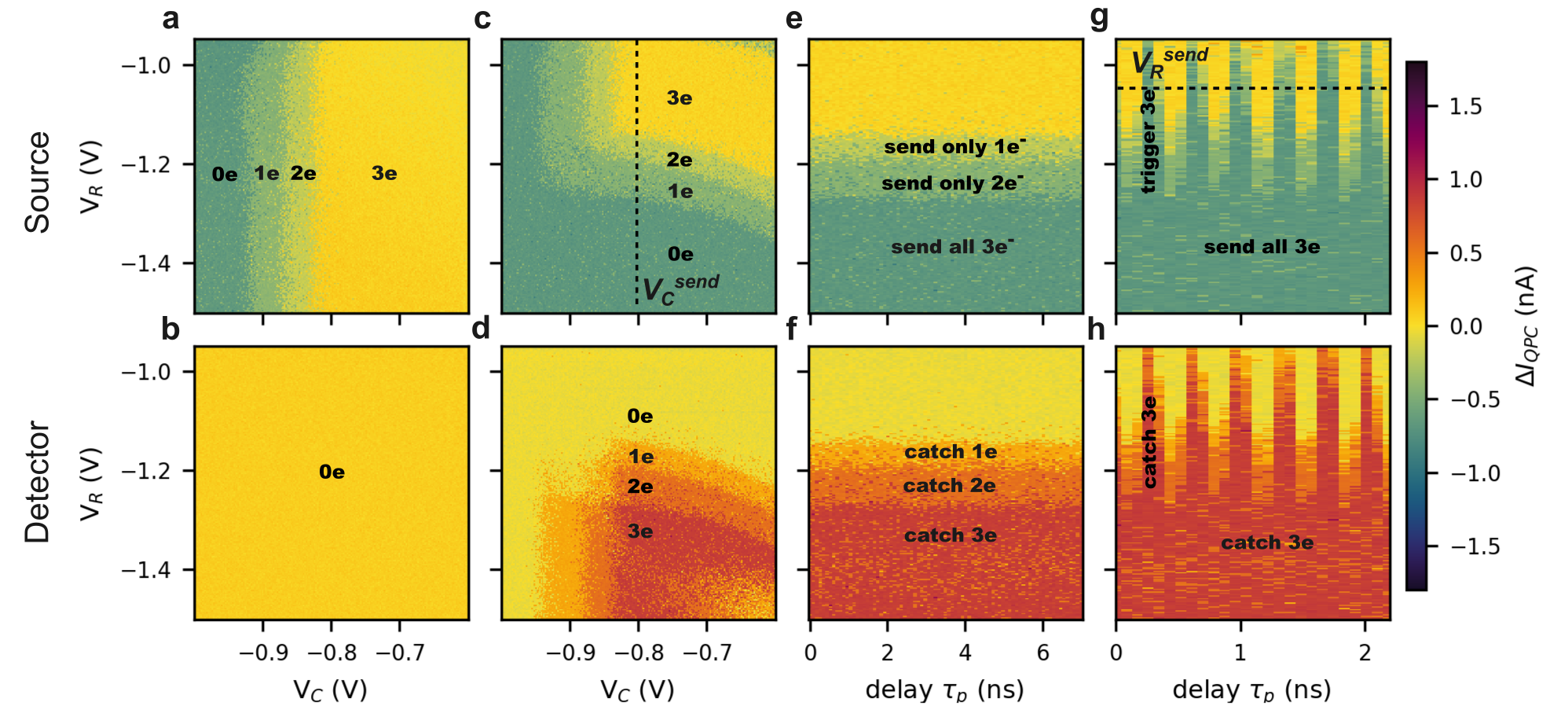}
    \caption{\textbf{Controlled sending of $N=3$ electrons.} 
    Top panels: Sending maps of the source S1. Bottom panels: Catching maps of the detector D2. The side gates of the central channel are adjusted to tilt the transverse potential well and direct all the electrons toward D2. 
    Three electrons have been initially loaded into the source QD. The voltage $V_\text{R}$ is then set to a more negative value, to prevent electrons from escaping the dot towards the reservoir when the SAW arrives. 
    \textbf{a,b}, Maps before applying the SAW, with all 3 electrons remaining in the source (yellow) and no electron in the detector (yellow). 
    \textbf{c,d}, Maps after applying the SAW, showing that 1, 2, or 3 electrons are transferred from the source (green) to the detector (orange). 
    \textbf{e,f}, Maps with the SAW applied, at a fixed voltage $V_\text{C}^\text{send}=\SI{-0.80}{\volt}$ (working point for sending), recorded without trigger pulse in a control experiment to check the electron sending as a function of $V_\text{R}$. 
    \textbf{g,h}, Maps with the SAW applied, and with a trigger pulse applied on the plunger gate (P) at different time delays $\tau_\text{P}$. Simultaneous sending of 3 electrons is achieved with a short negative pulse having $V_\text{P}=\SI{-0.50}{\volt}$ and $\delta t_\text{trig}=T_{\text{SAW}}/4\approx\SI{90}{\pico\second}$. The dotted line indicates the working point $V_\text{R}^\text{send}=\SI{-1.06}{\volt}$ used to trigger the sending of 3 electrons.
    }
    \label{figure_S2}
\end{figure}

Figure~\ref{figure_S2}a and \ref{figure_S2}b show the sending and catching maps of the source and detector QDs, when the SAW is not launched.
Note that the graph axis correspond to the gate voltages of the \textit{source} QD in both maps. 
These maps show that no electron is transferred, as expected in absence of SAW.

We then launch a \SI{180}{\micro\meter}-long SAW train (\SI{60}{\nano\second} of RF signal at \SI{3}{\giga\hertz}) and count the final number of electrons in the source and detector QDs to construct the sending and catching maps shown in Fig.~\ref{figure_S2}c and \ref{figure_S2}d. 
For gate voltages in the bottom-right sector of the maps, 1, 2, or 3 electrons have been transferred as indicated by the specific amplitude of the QPC current change. 
However, the electrons transported in this way are not transferred in the same potential minimum of the SAW train, but in random minima.

To place all the electrons into the same SAW minimum, the gate voltages $V_\text{C}^\text{send}$ and $V_\text{R}^\text{send}$ should be chosen in the top-right sector of the maps, such that the electrons remain trapped in the source QD even when the SAW train passes by.
The controlled sending is then achieved by applying an ultra-short negative voltage pulse to the plunger gate (P) of the source QD to rise the energy of the electrons and enable their catching and transport by the SAW, as shown in Fig.~\ref{figure_S2}g and \ref{figure_S2}h.
Applying an RF pulse with a time duration ($\delta t_\text{trig}$) shorter than the SAW period ensures that the electrons are placed in only one SAW minimum. 
Both the amplitude ($V_\text{P}$) and the time delay ($\tau_\text{P}$) of the trigger pulse are carefully adjusted to ensure that the three electrons are injected into the designated SAW minimum~\cite{takada2019sound, wang2023coulomb}. 
The values $V_\text{C}^\text{send}$ and $V_\text{R}^\text{send}$ are also optimized to get the highest transfer efficiency.

Since sending more than 3 electrons from a single source poses considerable technical challenges, sending 4 or 5 electrons is achieved by using the two sources S1 and S2 which are synchronised via their trigger pulse. 
For example, a droplet of 5 electrons is obtained by sending 3 electrons from S1 and 2 electrons from S2, all in the same SAW minimum. 
They are initially in different rails, but later join in the central channel.

\subsection{Placing all electrons in different SAW minima}

To investigate the partitioning of uncorrelated electrons, we have to place the electrons in different SAW minima. 
This configuration can be achieved by a combination of random and triggered electron sending, as well as a combination of the two electron sources S1 and S2. 
For the case of three electrons, we can for example proceed in the following ways. 

A first possibility is to load three electrons in a single source and simply apply a SAW train without triggering the plunger gate. 
In this case, $V_\text{R}^\text{send}$ is adjusted to allow the direct sending of the three electrons. 
When the sending takes place without trigger pulse, the electrons are automatically distributed over different and random SAW minima. 
This property is evidenced by the absence of correlation in the partitioning experiments performed in this way. 

A second possibility is to load three electrons in a single source, as above, but then send two electrons randomly and the third one with a trigger pulse.
In this case, $V_\text{R}^\text{send}$ is adjusted to allow the direct sending of only two electrons. 

A third possibility is to load two electrons in the source S1 and one electron in the source S2. 
We can either use the random sending procedure of the two electrons of S1, while the electron in S2 is sent using a trigger pulse at a delayed time near the end of the SAW train, or send the first electron of S1 randomly and the second electron of S1 with a trigger pulse, while the electron of S2 is sent with a trigger pulse at a later time. 

All these procedures ensure that the three electrons are in three different SAW minima.

\clearpage
\subsection{Detection of multiple electrons}

\begin{figure}[h]
    \centering   
    \includegraphics[scale=0.9]{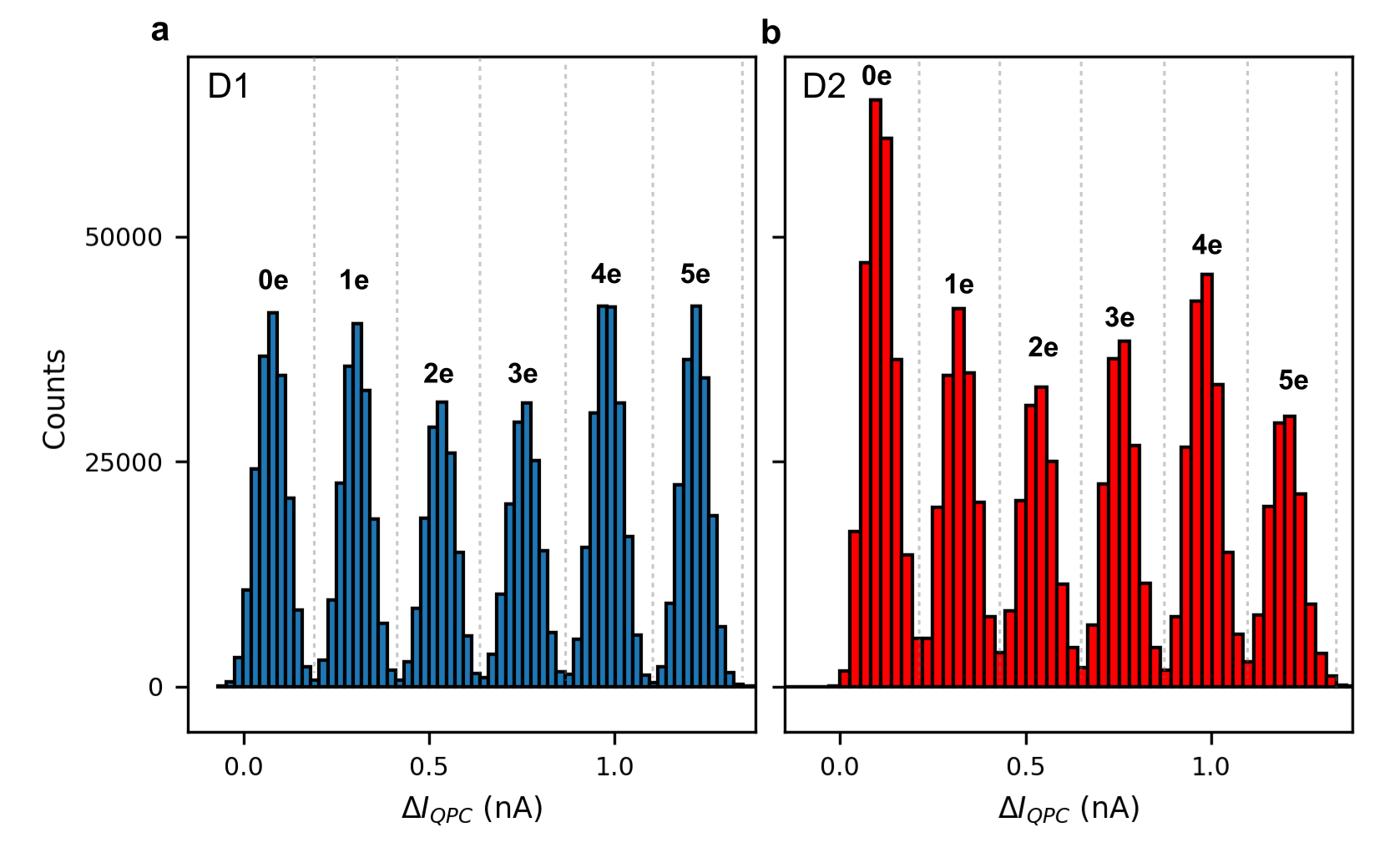}
    \caption{\textbf{Multi-electron detection.} 
    Histograms of the QPC current changes for (\textbf{a}) detector D1 and (\textbf{b}) detector D2, during the partitioning of $N=5$ electrons within the same SAW minimum (data collected for a full sweep of side-gate detuning before post-selection of residual transfer errors). This illustrates the capability to detect up to 5 interacting electrons.
    }
    \label{figure_S3}
\end{figure}

For partitioning experiments with $N$ electrons, the gate voltages of the detector QDs are carefully tuned to enable the catching of up to $N$ electrons. 
Figure~\ref{figure_S3} shows the histograms of current changes in the QPC sensors attached to the two detectors (D1 and D2) for the partitioning data of $N=5$ electrons in the same SAW minimum. 
The presence of well-separated peaks demonstrate the detector capability of catching up to 5 electrons.

\clearpage
\section{EFFECTIVE TEMPERATURE IN THE CENTRAL CHANNEL}
\label{supp:tuning}

In our experiment, the electron droplet is confined in three dimensions by the combined influence of the heterostructure band diagram (growth direction), the moving piezoelectric potential induced by the SAW (longitudinal direction) and the electrostatic potential induced by the surface gates (transverse direction). 
In the central channel, the presence of a narrow barrier gate in-between the two side gates introduces a small barrier in the middle of the transverse confinement potential.
In this section, we discuss the impact of the barrier-gate voltage $V_\text{B}$ on electron partitioning at the Y-junction, and use this adjustable parameter to estimate the effective electron temperature with a model of thermally activated hopping in a double-well potential.

\subsection{Partitioning versus barrier-gate voltage for one electron}

\begin{figure}[h]
    \centering
    \includegraphics[scale=0.9]{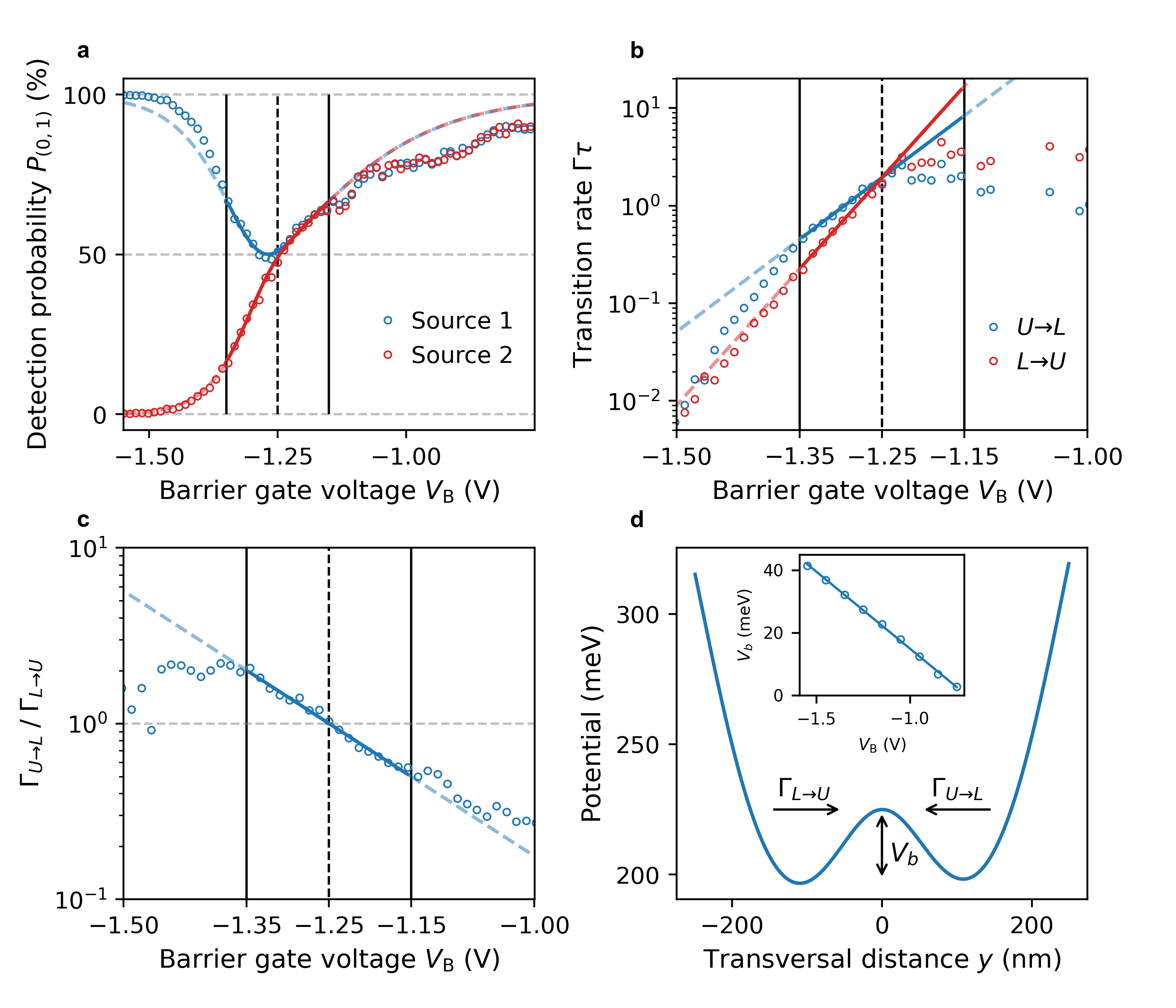}
    \caption{\textbf{Transition rates across the central-channel barrier.} 
    \textbf{a}, Detection probabilities for single-electron partitioning as function of the barrier-gate voltage $V_\text{B}$ and for fixed side-gate voltages $V_\text{U}=\SI{-1.108}{\volt}$ and $V_\text{L}=\SI{-1.098}{\volt}$. The electron is sent either from the source S1 or from the source S2. Blue and red solid lines correspond to the two-site model fitted to the experimental data within the fit window around the working point at $V_\text{B}^0=\SI{-1.25}{\volt}$, while dashed lines show the model outside the fit window (indicated by the vertical lines). 
    \textbf{b}, Transition rates calculated from the experimentally measured detection probabilities after fitting with the two-site model. 
    \textbf{c}, Ratio of transition rates for the two directions. 
    \textbf{d}, Electrostatic simulation of the potential energy along the transverse direction at $V_\text{B}^0$. The inset shows the barrier-height dependence on barrier-gate voltage.
    }
    \label{figure_S4}
\end{figure}

With side-gate voltages held constant around zero detuning to have symmetric partitioning, we measured the partitioning probability as function of the barrier-gate voltage $V_\text{B}$ for a single electron sent from source S1 or from source S2 (Fig.~\ref{figure_S4}a). 
At $V_\text{B}=\SI{-1.5}{\volt}$, the upper and lower rails are fully isolated, preventing electron from transitioning from one side to the other. 
For $V_\text{B}\ge\SI{-1.25}{\volt}$, electrons sent from sources S1 and S2 show the same partitioning result $P_{(0,1)}^{\rm S1}=P_{(0,1)}^{\rm S2}$ and hence are statistically indistinguishable. 
This property is the consequence of the very long (\SI{40}{\micro\meter}) central channel which gives the electron enough time (\SI{14}{\nano\second}) to equilibrate between the two rails by tunneling or hopping through the barrier.

For all partitioning experiments reported in this work, the barrier-gate voltage has been fixed at $V_\text{B}^0=\SI{-1.25}{\volt}$, such that both sources S1 and S2 can be used indifferently, with $P_{(N-n,\,n)}^{\rm S1} = P_{(N-n,\,n)}^{\rm S2}$, simply noted $P_{(N-n,\,n)}$ (for supporting data, see Supplementary Note~\ref{supp:indistinguishability}).

\subsection{Extraction of transition rates with two-site master equations}
\label{supp:two_site_model}

To analyse the single-electron partitioning data, the double-well potential of the central channel is first modelled by a simplified two-site system, which will be refined later with a 1D quartic potential.

We model the single-electron partitioning data with the two-site master equations 
\begin{align}
    \partial_t 
    \begin{pmatrix}
        P_\text{U} \\ P_\text{L}
    \end{pmatrix} = 
    \begin{pmatrix}
        -\Gamma_{\text{U}\to\text{L}} &  \Gamma_{\text{L}\to\text{U}} \\
         \Gamma_{\text{U}\to\text{L}} & -\Gamma_{\text{L}\to\text{U}}
    \end{pmatrix}
    \begin{pmatrix}
        P_\text{U} \\ P_\text{L}
    \end{pmatrix}, \label{eq:two_site_master}
\end{align}
where the time-independent transition rates $\Gamma_{\text{U}\to\text{L}}$ and $\Gamma_{\text{L}\to\text{U}}$ correspond to stochastic single-electron transitions from the upper to the lower rail (and vice-versa). 
Solving this system of coupled linear equations, two solutions for $P_\text{U}(t)$ are obtained corresponding to two different initial conditions: electron sent from S1 ($P_\text{U}^\text{S1}(0)=1$) and electron sent from S2 ($P_\text{U}^\text{S2}(0)=0$). 
The experimentally measured probabilities $P_{(0,1)}$ correspond to these solutions at $t=\tau$, where $\tau$ is the duration of the electron flight along the central channel,
\begin{align}
\begin{split}
    P_\text{U}^\text{S1}(\tau) &= \frac{\Gamma_{\text{L}\to\text{U}} + \Gamma_{\text{U}\to\text{L}} \, e^{-(\Gamma_{\text{U}\to\text{L}} + \Gamma_{\text{L}\to\text{U}})\tau}}{\Gamma_{\text{U}\to\text{L}} + \Gamma_{\text{L}\to\text{U}}}, \\
    P_\text{U}^\text{S2}(\tau) &= \frac{\Gamma_{\text{L}\to\text{U}} - \Gamma_{\text{L}\to\text{U}} \, e^{-(\Gamma_{\text{U}\to\text{L}} + \Gamma_{\text{L}\to\text{U}})\tau}}{\Gamma_{\text{U}\to\text{L}} + \Gamma_{\text{L}\to\text{U}}}.
    \label{eq:two_site_sol}
\end{split}
\end{align}

By inverting these relations, the transition rates $\Gamma_{\text{L(U)}\to\text{U(L)}}$ can be expressed in terms of the experimentally measured probabilities, as shown in Fig.~\ref{figure_S4}b. 
In the vicinity of the working point $V_\text{B}^0$ selected for the partitioning experiments, the transition rates can be approximated as exponentially dependent on the barrier-gate voltage. 
For voltages above $V_\text{B}^0$, the characteristic transition times $1/\Gamma_{\text{L(U)}\to\text{U(L)}}$ are much shorter than the duration of the flight in the central channel, giving rise to statistical equilibrium between the two rails. 
The two electron exchange rates are not equal, and their ratio plotted in Fig.~\ref{figure_S4}c is exponentially dependent on $V_\text{B}$.

Based on these observations, we express the rates as
\begin{align}
\begin{split}
    \Gamma_{\text{U}\to\text{L}} &= \Gamma_0 \, e^{c_{\text{B}}^{\text{U}\to\text{L}}(V_\text{B} - V_\text{B}^0)} \\
    \Gamma_{\text{L}\to\text{U}} &= \Gamma_0 \, e^{c_{\text{B}}^{\text{L}\to\text{U}}(V_\text{B} - V_\text{B}^0)}
    \label{eq:two_site_gammas}
\end{split}
\end{align}
and fit the measured probabilities with the solutions \eqref{eq:two_site_sol} together with the expressions \eqref{eq:two_site_gammas} as shown by solid lines in Fig.~\ref{figure_S4}a. 
The model agrees well with experimental data within the fitting window and predicts qualitatively the measured probabilities outside the window.
The best fit parameters are $c_{\text{B}}^{\text{U}\to\text{L}}=\SI{14.6}{\per\volt}$, $c_{\text{B}}^{\text{L}\to\text{U}}=\SI{21.5}{\per\volt}$ and $\Gamma_0\tau = 1.94$. 
The difference between the two slope coefficients $c_\text{B}$ corresponds to an asymmetric influence of the barrier gate on the electrostatic potential of the two sites.
The inferred value of $\Gamma_0\tau$ implies that at the working point $V_{\text{B}}^{0}$ the single-electron probabilities have equilibrated to within $2\%$.

\subsection{Thermally-activated hopping in a quartic double-well potential}
\label{supp:quartic_potential}

The above barrier-gate dependence of the transition rates can be used to extract quantitative information about the energy properties of the electron in the central channel.
For this purpose, we consider a simple 1D double-well potential in form of a quartic polynomial, keeping the conventional notation~\cite{weiss1983complex},
\begin{align}
    V(y) = V_b + \mu_q \frac{y}{y_0} - 8V_b \frac{y^2}{y_0^2} + 16V_b \frac{y^4}{y_0^4}, \label{eq:quartic_pot}
\end{align}
where $V_b$ is the energy barrier (controlled by the barrier-gate voltage $V_\text{B}$) and $\mu_q$ is the energy detuning (controlled by the side-gates voltage difference $\Delta=V_\text{U}-V_\text{L}$). 
Tuned to symmetry, the potential has two minima at $y=\pm y_0/2$ which we identify with the two sites of the kinetic model considered in previous section. 
Note that the same potential is used also for multi-electron simulations described in the Methods.

The theory of transition rates \cite{weiss2012quantum, hanggi1990reaction} generically predicts two regimes that depend on the temperature $T$: tunneling for $T<T_0$ and hopping for $T>T_0$, where $T_0$ is the tunneling-to-hopping crossover temperature determined by the curvature of $V(y)$ at the top of the barrier.
For the quartic potential, it is given by $k_BT_0=3\hbar\omega_y/16$ in terms of the curvature $\omega_y=\sqrt{32V_b/(my_0^2)}$ at the minima of the double-well potential ($m=0.067m_e$ for electrons in GaAs).

In the regime of thermally-activated hopping, but still in the high barrier limit $\beta V_b \gg 1$ where $\beta=1/k_\text{B}T$, the temperature-dependent hopping rates (for weak detuning $\mu_q$) can be approximated as \cite{affleck1981quantum}
\begin{align}
\begin{split}
    \Gamma_{\text{U}\to\text{L}} &= \Gamma_0^\text{(cl)} \, e^{-\beta (V_b-\mu_q/2)}, \\
    \Gamma_{\text{L}\to\text{U}} &= \Gamma_0^\text{(cl)} \, e^{-\beta (V_b+\mu_q/2)}.
    \label{eq:1D_gammas}
\end{split}
\end{align}
where the prefactor $\Gamma_0^\text{(cl)}$ is the classical attempt rate for hopping.
These relations can be used to extract the effective electron temperature $T$ in the central channel from the barrier-gate dependence of the experimental transition rates obtained in the previous section.
However, to determine the effective temperature in Kelvin, we need the conversion factor between the barrier-gate voltage $V_\text{B}$ and the barrier energy $V_b$.

We estimate the numerical values of the parameters of the quartic potential Eq.~\eqref{eq:quartic_pot} using NextNano~\cite{birner2007nextnano} self-consistent 3D simulations of the electrostatic potential induced by the surface gates in the modulation-doped GaAs/AlGaAs heterostructure, including non-linear screening by the 2DEG in non-depleted regions. 
For the actual values of voltages applied to the side gates and barrier gate, the central channel is depleted and forms a double-well potential (Fig.~\ref{figure_S4}d). 
The potential profile has been simulated for a range of experimental voltages $V_\text{B}$ in the symmetric configuration $V_\text{U}= V_\text{L}$ corresponding to $\mu_q=0$. 
The energy barrier height $V_b$ (energy difference between the central maximum and the lateral minima) is shown in the inset as function of $V_\text{B}$.
At the working point $V_\text{B}^0=\SI{-1.25}{\volt}$, the barrier height is $V_b=\SI{27.5}{\milli\electronvolt}$ and the distance between the two minima is $y_0=\SI{220}{\nano\meter}$, which gives a tunneling-to-hopping crossover temperature $k_\text{B}T_0=\SI{0.85}{\milli\electronvolt}$ (about \SI{10}{\kelvin}). 
The inset also indicates a linear relationship between the experimental voltage $V_\text{B}$ and the energy barrier height $V_b$ expressed as
\begin{align}
    V_b = \alpha_{\text{B}} (V_\text{B} - V_\text{B,flat}) 
    \label{eq:barrier_gate_rel}
\end{align}
with a lever-arm factor $\alpha_{\text{B}}=\SI{-49.2}{\milli\electronvolt\per\volt}$ and a flat-well voltage $V_\text{B,flat}=\SI{-0.70}{\volt}$.
Thanks to this linear dependence, the exponents of the hopping rates in Eq.~\eqref{eq:1D_gammas} can be compared with those of Eq.~\eqref{eq:two_site_gammas} to get the relation 
\begin{align}
    \beta|\alpha_\text{B}| = \bar{c}_\text{B} 
\end{align}
where the coefficient $\bar{c}_\text{B}=(c_\text{B}^{\text{U}\to\text{L}}+c_\text{B}^{\text{L}\to\text{U}})/2=\SI{18}{\per\volt}$ is the average of the two fitted coefficients. 
The effective electron temperature in the moving confinement potential is then obtained as 
\begin{align}
    k_\text{B}T \equiv \beta^{-1} = \frac{|\alpha_\text{B}|}{\bar{c}_\text{B}} = \SI{2.73}{\milli\electronvolt}  
    \text{ (\SI{32}{\kelvin})}  \, .
\end{align}
The value $T>T_0$ confirms the consistency of our assumptions for Eq.~\eqref{eq:1D_gammas} that electron exchanges between the two wells takes place in the classical hopping regime.

\subsection{Discussion of the effective temperature}
\label{supp:temperature_discussion}

The electrostatic simulation also provides the detuning lever arm factor $\alpha=\SI{0.16}{\milli\electronvolt\per\milli\volt}$, which is introduced in the main text to relate the energy detuning $\mu$ of the Ising model to the side-gates voltage difference by $\mu=-\alpha(\Delta-\Delta_0)$.
The global fitting parameter $\alpha/k_\text{B}T=\SI{0.064}{\per\milli\volt}$ obtained by fitting the multi-electron partitioning data with the Ising model (see Extended Data Table 1) can therefore provide another estimate of the effective electron temperature, and we get $k_\text{B}T=\SI{2.5}{\milli\electronvolt}$ ($\SI{29}{\kelvin}$) in good agreement with the previous value.
The interaction strength $U$ of the Ising model can be then deduced from the second fitting parameter $U/k_\text{B}T$, and we find that $U$ decreases from $3.8$ to $\SI{2.2}{\milli\electronvolt}$ when $N$ increases from 2 to 5.

The values of the effective temperature estimated above are in reasonable agreement with $T=\SI{25}{\kelvin}$ found from matching the observed multi-electron correlations to Monte-Carlo simulations, as described in the main text.
The latter also gives an independent way to estimate the level-arm factor $\alpha$ resulting in values within the range of $0.13$ to $\SI{0.15}{\milli\electronvolt\per\milli\volt}$, consistent with $\SI{0.16}{\milli\electronvolt\per\milli\volt}$ estimated from electrostatic simulations. 

Nevertheless, one has to exercise caution in interpreting the hopping-model estimates, since they only include single-electron hopping between two sites with a constant effective rate during the propagation time $\tau$ before reaching the Y-junction.
The two-site hopping model averages over the disorder in the central channel and does not take into account the excitation of the electrons as they move along the not perfectly uniform potential landscape~\cite{takada2019sound}.
We attribute the high effective temperature of the electron droplet to this physical effect.
Additionally, non-equilibrium dynamics of electrons on a scale much shorter than $\tau$ during partitioning \cite{fricke2013counting} at the exit Y-junction may also contribute to the freeze-out temperature estimated in the sudden quench approximation.

\clearpage
\section{STATISTICAL INDISTINGUISHABILITY OF S1 AND S2 ELECTRONS} 
\label{supp:indistinguishability}

To prepare an electron droplet with $N>3$, we utilise both the S1 and S2 sources and synchronise them to place all the electrons into the same SAW minimum.
However, we must ensure that the electrons injected from sources S1 and S2 are statistically indistinguishable. 
This is accomplished by first tuning the barrier height of the central channel to achieve indistinguishability for $N=1$ electron, and then checking the indistinguishability in the partitioning experiments for $N=2$ to 4 electrons.

\subsection{Selection of a suitable barrier-gate voltage}

\begin{figure}[h]
    \centering
    \includegraphics[scale=0.9]{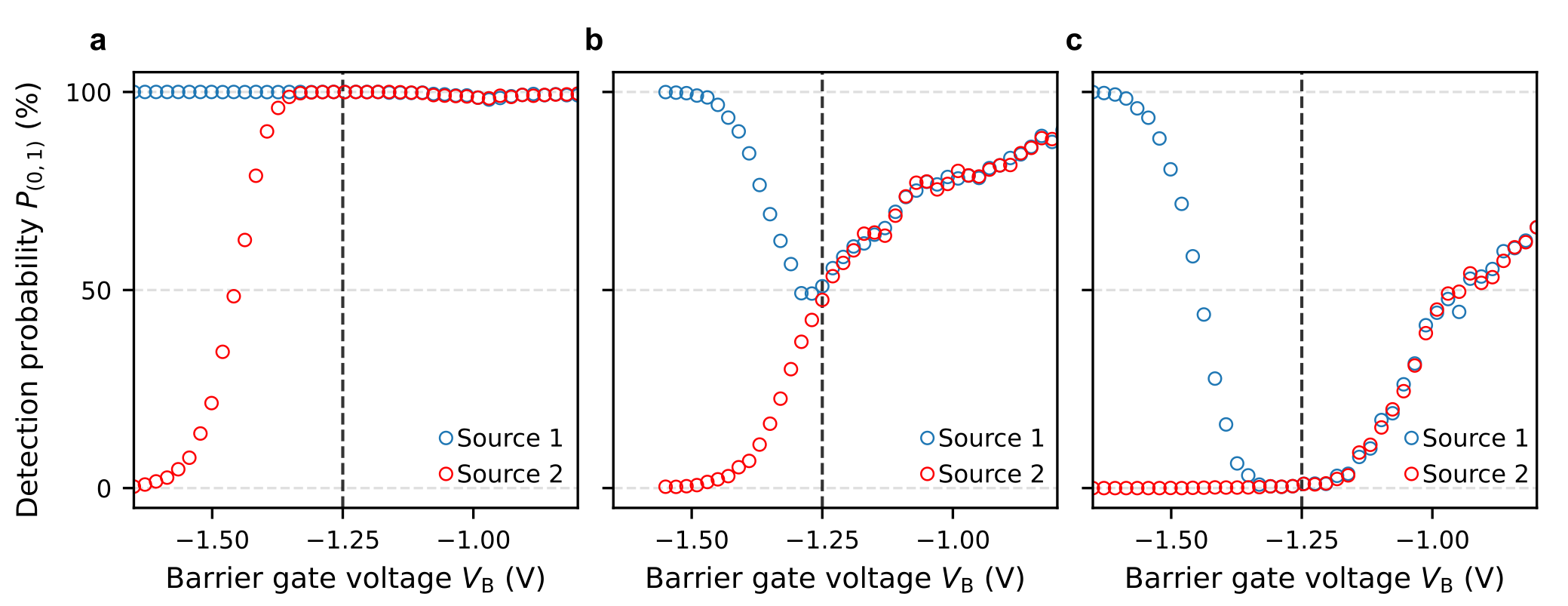}
    \caption{
    \textbf{Barrier-gate dependence and electron indistinguishability.} 
    Partitioning probability of a single electron as function of the barrier-gate voltage $V_\text{B}$. The blue circles represent injection from the upper source S1, while the red circles indicate injection from the lower source S2. Each data point is based on 3,000 single-shot measurements and represents the probability $P_{(0,1)}$ to detect the electron in the upper detector D1. The vertical dashed line indicates the operating point $V_\text{B}^0=\SI{-1.25}{\volt}$.
    \textbf{a}, The side gates are set to the large positive detuning $\Delta=\SI{85}{\milli\volt}$. 
    \textbf{b}, The side gates are set to the near-zero detuning $\Delta=\SI{-10}{\milli\volt}$ where a symmetric partitioning is achieved at $V_\text{B}^0$.
    \textbf{c}, The side gates are set to the large negative detuning $\Delta=\SI{-85}{\milli\volt}$.
    }
    \label{figure_S5}
\end{figure}

A single electron is injected from either source S1 or source S2, and the partitioning probability is recorded as a function of the barrier-gate voltage $V_\text{B}$, with the side gates kept fixed at three different detuning values, as shown in Fig.~\ref{figure_S5}.

For $V_\text{B}<\SI{-1.50}{\volt}$ in panel b (near zero detuning) or $V_\text{B}<\SI{-1.60}{\volt}$ in panels a and c (for large detuning), a high barrier prevents the electron from hopping between the two wells of the transverse double-well potential, such that the two wells are fully isolated. 

For $V_\text{B}\geq\SI{-1.25}{\volt}$, the partitioning probabilities are identical for electron injections from source S1 and from source S2.
The operating point $V_\text{B}^0=\SI{-1.25}{\volt}$ was selected to have electron indistinguishability, while maintaining a strong confinement within the central channel. 
At this barrier-gate voltage, an electron injected from source S1 is statistically equivalent to an electron injected from source S2, for all the detuning voltages used in the partitioning experiments.

\subsection{Comparison of different loading configurations}

\begin{figure}[h]
    \centering
    \includegraphics[scale=0.9]{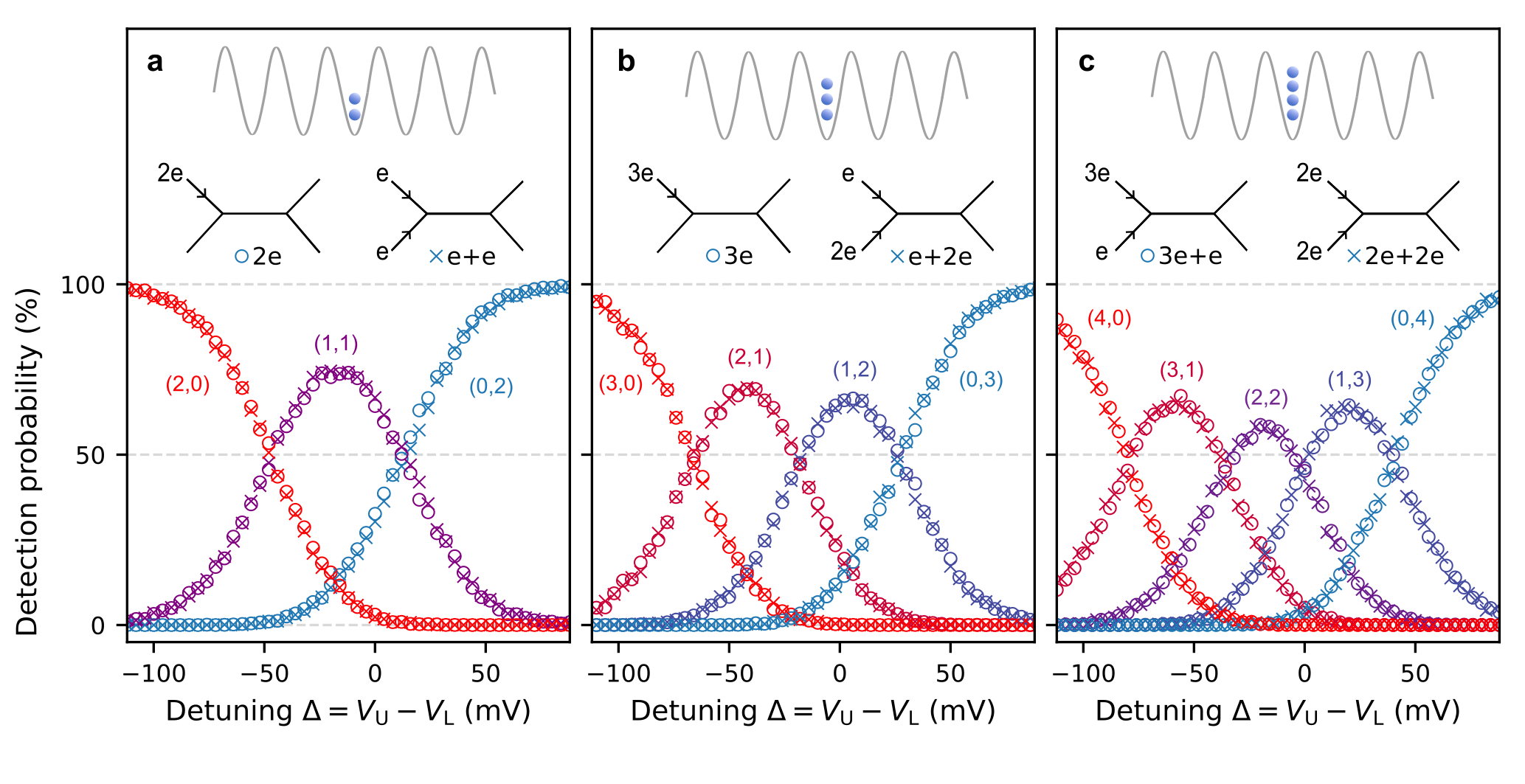}    
    \caption{
    \textbf{Partitioning of electron droplets with different loading configurations.}
    The barrier-gate voltage is set at the same working point $V_\text{B}^0$ as in the experiments of the main text. The illustration at the top of each panel depicts the loading configurations. The two sources S1 and S2 are synchronised to load all electrons into the same SAW minimum. Each data point is based on 3,000 single-shot measurements.
    \textbf{a}, Partitioning probabilities for an $N=2$ electron droplet. The configurations are 2e (two electrons from source S1) and e+e (one electron from source S1 and one from source S2).
    \textbf{b}, Partitioning probabilities for an $N=3$ electron droplet. The configurations are 3e (three electrons from source S1) and e+2e (one electron from source S1 and two from source S2).  
    \textbf{c}, Partitioning probabilities for an $N=4$ electron droplet. The configurations are 3e+e (three electrons from source S1 and one from source S2) and 2e+2e (two electrons from source S1 and two from source S2). 
    }
    \label{figure_S6}
\end{figure}

Here we compare the partitioning of multi-electron droplets created from different configurations of electron number in sources S1 and S2. 
Fixing the barrier-gate voltage at the operating point $V_\text{B}^0$, we record the partitioning probabilities of droplets with $N=2$ to 4 electrons, using different distributions of the $N$ electrons between the two sources.
Figure~\ref{figure_S6} shows that the probabilities are identical regardless of the electron loading configuration.
This observation experimentally confirms that, during their evolution along the central channel, the electrons of the droplet have lost information about their source, becoming statistically indistinguishable.

\clearpage
\section{RECONSTRUCTION FORMULA FOR INDEPENDENT ELECTRONS} 
\label{supp:reconstruction}

\begin{figure}[h]
    \centering
    \includegraphics[scale=0.9]{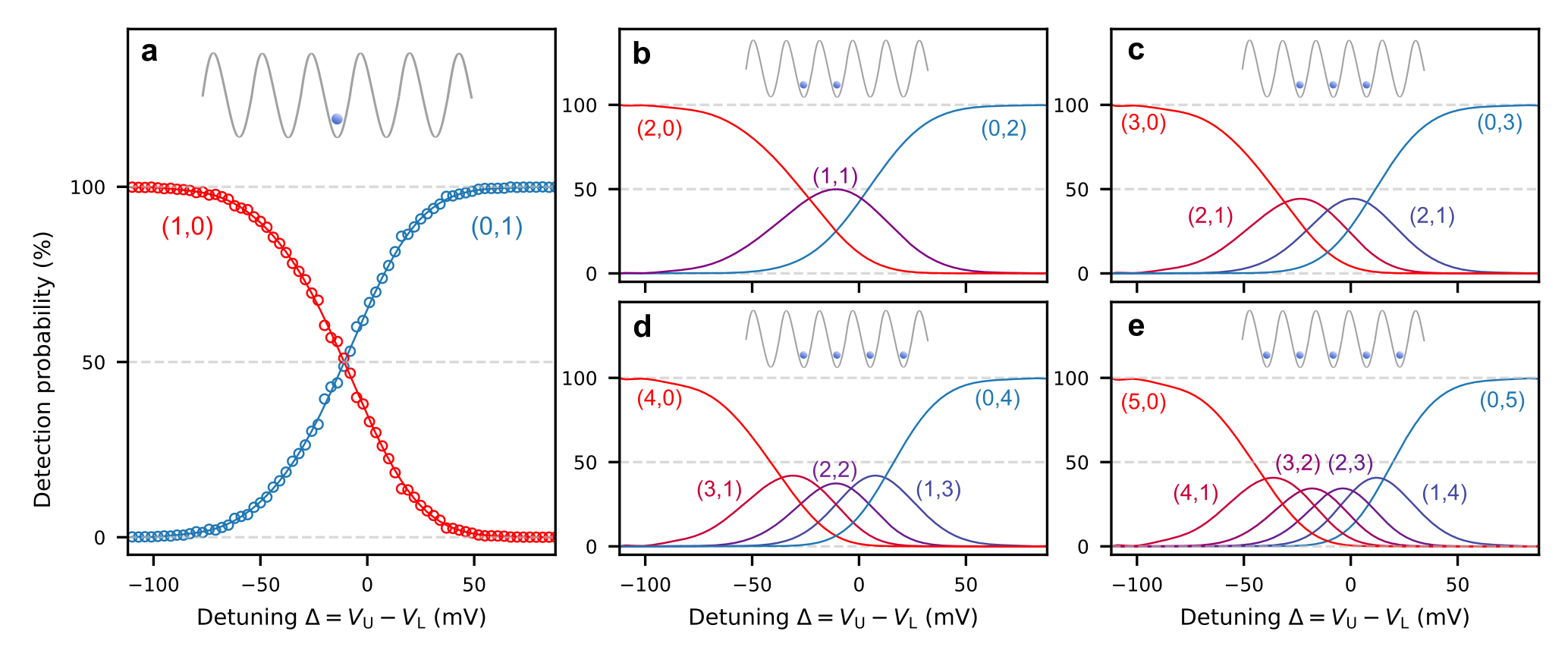}
    \caption{\textbf{Reconstruction of partitioning probabilities of uncorrelated electrons using single-electron data.} 
    \textbf{a} Probability distributions for single-electron partitioning as function of side-gates detuning, for $V_\text{B}=\SI{-1.25}{\volt}$. Each data point is based on 3,000 single-shot measurements. Lines are $15^\text{th}$-order polynomial fits, which are used to model multi-electron partitioning probabilities corresponding to uncorrelated electrons distributed among different SAW minima. 
    \textbf{b-e}, Reconstruction of the partitioning probabilities for $N=2$, 3, 4, and 5 uncorrelated electrons placed in different SAW minima, using the expressions given in Table~\ref{tab:equations}.}
    \label{figure_S7}
\end{figure}

Table \ref{tab:equations} explains how to reconstruct the partitioning probabilities of $N>1$ independent electrons from the knowledge of the $N=1$ partitioning probabilities. This reconstruction is illustrated in Fig.~\ref{figure_S7}.

\begin{table}[h]
    \centering
    \begin{tabular}{c|c|c|c}
        \textbf{e/e} & \textbf{e/e/e} & \textbf{e/e/e/e} & \textbf{e/e/e/e/e} \\
        \hline
        \begin{tabular}{l}
            $ P_{(2,0)} = P_{(1,0)}^{2} $ \\
            $ P_{(1,1)} = 2P_{(1,0)}P_{(0,1)} $ \\
            $ P_{(0,2)} = P_{(0,1)}^{2} $
        \end{tabular} &
        \begin{tabular}{l}
            $ P_{(3,0)} = P_{(1,0)}^{3} $ \\
            $ P_{(2,1)} = 3P_{(1,0)}^{2}P_{(0,1)} $ \\
            $ P_{(1,2)} = 3P_{(1,0)}P_{(0,1)}^{2} $ \\
            $ P_{(0,3)} = P_{(0,1)}^{3} $
        \end{tabular} &
        \begin{tabular}{l}
            $ P_{(4,0)} = P_{(1,0)}^{4} $ \\
            $ P_{(3,1)} = 4P_{(1,0)}^{3}P_{(0,1)} $ \\
            $ P_{(2,2)} = 6P_{(1,0)}^{2}P_{(0,1)}^{2} $ \\
            $ P_{(1,3)} = 4P_{(1,0)}P_{(0,1)}^{3} $ \\
            $ P_{(0,4)} = P_{(0,1)}^{4} $
        \end{tabular} &
        \begin{tabular}{l}
            $ P_{(5,0)} = P_{(1,0)}^{5} $ \\
            $ P_{(4,1)} = 5P_{(1,0)}^{4}P_{(0,1)} $ \\
            $ P_{(3,2)} = 10P_{(1,0)}^{3}P_{(0,1)}^{2} $ \\
            $ P_{(2,3)} = 10P_{(1,0)}^{2}P_{(0,1)}^{3} $ \\
            $ P_{(1,4)} = 5P_{(1,0)}P_{(0,1)}^{4} $ \\
            $ P_{(0,5)} = P_{(0,1)}^{5} $
        \end{tabular} \\
    \end{tabular}
    \caption{Expressions for reconstructing the partitioning probabilities $P_{(N-n,\,n)}$ of $N=2$, 3, 4, and 5 non-interacting electrons placed in different SAW minima, using the experimental partitioning probabilities $P_{(1,0)}$ and $P_{(0,1)}$ of a single electron ($N=1$), as illustrated in Fig.~\ref{figure_S7}.}
    \label{tab:equations}
\end{table}

\clearpage
\section{COUNTING STATISTICS AND CUMULANTS} 
\label{supp:cumulantsGeneral}

\subsection{Multivariate and univariate cumulants}

The outcome of the partitioning experiment can be represented by the probabilities $p_n$ to detect $n$ electrons at a chosen detector (here D1), as done in full counting statistics (FCS) \cite{levitov1996electron}. This probability distribution can be characterised in terms of either moments $\langle n^k \rangle$ or cumulants $\llangle n^k \rrangle$, which are related to each other by the formal expansion of the appropriate generating function 
\begin{align}
    \ln \langle e^{z n} \rangle = \ln\left( \sum_{k=0}^{\infty} \langle n^k \rangle \frac{z^k}{k!} \right) = \sum_{k=1}^{\infty} \llangle n^k \rrangle \frac{z^k}{k!} 
    \label{eq:generatecum}
\end{align}
where $\langle f(n) \rangle = \sum_n p_n f(n)$ denotes averaging over the probabilities of the FCS.
One can express moments through cumulants
\begin{align}
    \langle n^k \rangle = \sum_{j=0}^k B_{kj} \left ( \llangle n \rrangle,\llangle n^2 \rrangle, \ldots, \llangle n^{j} \rrangle
    \right )
    \label{eq:nkviankcum}
\end{align}
and vice versa
\begin{align}
   \llangle n^k \rrangle = \sum_{j=1}^k
    (j - 1)! (-1)^{j - 1} B_{kj}
    \left ( \langle n \rangle,\langle n^2 \rangle, \ldots, \langle n^{k-j+1} \rangle
    \right ),
    \label{eq:nkcumviank}
\end{align}
where $B_{kj}$ are the partial Bell polynomials~\cite{comtetbook}.

Two properties of regular (univariate) cumulants are noteworthy in the context of studying correlations between partitioning events, for an ensemble of $N$ particles. Firstly, that they are additive for independent variables. In our case, the number $n$ of particles detected at D1 can be written as the sum $n=\sum_{j=1}^{N}T_j$ of binary variables $T_j\in\{0,1\}$ coding the absence (0) or presence (1) of the particle $j$ in the detector D1 for a given realization of the partitioning experiment. If the variables $T_j$ are statistically independent ($p_n$ following a binomial distribution), then $\llangle n^k \rrangle_N=N \llangle n^k \rrangle_1$. Secondly, if $p_n$ follows Gaussian distribution, $\llangle n^k \rrangle=0$ if $k>2$. Note that this property is valid only for continuous $n$, so higher-order ($k>2$) non-zero cumulants serve as an indication of deviations from Gaussian behaviour but do not distinguish between the effects of discrete $n$, finite $N$, or  correlations induced by interactions. 

The univariate cumulants $\llangle n^k \rrangle$ discussed above are therefore not good indicators of inter-particle correlations, as they remain non-zero even when the partitioning is independent. Instead, we consider multivariate cumulants $\llangle T_{1}^{i_1} \dots T_{N}^{i_N} \rrangle$ defined with a generating function
\begin{align}
    \ln{\langle e^{\mathbf{z}\cdot \mathbf{T}} \rangle} = \sum\limits_{i_1, i_2, \dots, i_N} \llangle T_{1}^{i_1} \dots T_{N}^{i_N} \rrangle \frac{z_{1}^{i_1}\dots z_{N}^{i_N}}{i_1 ! \dots i_N !}
    \end{align}
where $\mathbf{T}=(T_1, \dots, T_N)$ and $\mathbf{z}=(z_1, \dots, z_N)$. The sum $i_1+i_2+ \ldots +i_N=k$ counts the number of variables $T_j$ (including possible repetitions) involved in the cumulant and represents the order of the cumulant. For a given $k$, there is one multivariate cumulant for each $\{i_1,...,i_N\}$ leading to a large number of non-equivalent multivariate cumulants if the variables are not equivalent. 

Multivariate cumulants carry more information than univariate cumulants because they capture correlations between multiple variables, allowing for the analysis of many-body systems with correlations between several particles. The key advantage is that a multivariate cumulant between two independent variables $T_a$ and $T_b$ is zero, hence higher-order cumulants reveal the presence of multi-body correlations. In statistics, the multivariate cumulants~\cite{Indian1962} are also referred to as irreducible correlators, and in field theory, they are known as connected diagrams or connected correlators. Multivariate cumulants and moments are related by general formulas; this relationship is a combinatorial procedure that involves summing over all possible partitions of the indices corresponding to the order $k$ of the cumulant~\cite{Gardiner1986HandbookOS}. A more detailed example of applying the general formulas for specific symmetry (determined by electron configuration in SAW minima) is provided in Supplementary Note \ref{supp:2e2epartitioning}.

\subsection{Relation between multivariate cumulants and full counting statistics}

FCS measures the cumulative observable $n=\sum_j T_j$ which is symmetric under the permutation of particles. Here we show that knowing the FCS is equivalent to the knowledge of all symmetrised multivariate moments $m_k$, as expressed by Eq.~\eqref{eq:mpviap} in Methods. For permutation-symmetric distributions, any multivariate moment or cumulant of $k$ distinct variables depends only on $k$, hence in the fully symmetric case, the knowledge of $m_k$ implies the knowledge of the corresponding multivariate cumulants $\kappa_k$, which can therefore be calculated from the probabilities $p_n$ of the FCS.

We define the fully symmetrised $k$-th order multivariate moments $m_k$ as an average over all distinct combinations of $k$ variables out of $N$
\begin{align} \label{eq:mkdefsupp}
    m_k = \binom{N}{k}^{-1} \langle e_k(T_1,\ldots,T_N) \rangle
\end{align}
where $\binom{N}{k}=\frac{N!}{k!(N-k)!}$ is the binomial coefficient and 
\begin{align} 
    e_k(T_1,\ldots,T_N) = \sum_{1 \leq j_1 < j_2 < \dots < j_k \leq N} T_{j_1}T_{j_2}\ldots T_{j_k}
\end{align}
is the elementary symmetric polynomial in the $N$ variables $T_1,\ldots,T_N$, which is made of $\binom{N}{k}$ terms corresponding to all possible combinations of $k$ variables $T_j$ out of $N$.

Since $T_{j_1}T_{j_2}\ldots T_{j_k}$ equals to 1 if and only if the subset $\{j_1,j_2,\ldots,j_k\}$ of $k$ variables is among the ensemble of particles detected in D1, the quantity $\langle e_k(T_1\,\ldots,T_N) \rangle$ is the sum of all the probabilities $p_n$ with $n \ge k$ multiplied by the number $\binom{n}{k}$ of possibilities to take $k$ terms out of $n$. In terms of FCS, the symmetrised multivariate moments can therefore be written 
\begin{align}
   m_k = \binom{N}{k}^{-1} \sum_{n=k}^N \binom{n}{k} p_n 
   \label{eq:mpviapSupp}.
\end{align}

This result can be mathematically demonstrated as follows. In our case, the variables $T_j$ are binary ($T_j \in \mathbb{Z}^2$) and idempotent ($T_j^2 = T_j$). Under these conditions, a multiplication law $e_1 \, e_j = j \, e_j+ (j+1) \, e_{j+1}$ holds true. Using this multiplication law, one can prove the identity $e_k=\binom{e_1}{k}$ by induction. As $e_1=\sum_{j=1}^{N} T_j=n$, the elementary symmetric polynomial $e_k$ is simply equal to the binomial coefficient $\binom{n}{k}$. The symmetrised multivariate moment defined as $m_k = \binom{N}{k}^{-1} \langle e_k \rangle$ writes
\begin{align}
   m_k = \binom{N}{k}^{-1} \left\langle \binom{n}{k} \right\rangle = \binom{N}{k}^{-1} \sum_{n=k}^N \binom{n}{k} p_n 
   \label{eq:mpviap2}
\end{align}
as anticipated above, the average being replaced by its definition in terms of probabilities. We also obtain the relation
\begin{align}
    m_k = \frac{\langle(n)_k\rangle}{(N)_k} 
    \label{eq:vaifactorialcumulant}
\end{align}
where $(n)_k=n(n-1)\ldots(n-k+1)$ is the falling factorial and $\langle(n)_k\rangle$ is called the factorial moment of order $k$. 

Similar to Eq.~\eqref{eq:mkdefsupp}, we define the symmetrised $k$-th order multivariate cumulants
\begin{align}
    \kappa_k = \binom{N}{k}^{-1} \llangle e_k(T_1 \, \ldots T_N) \rrangle 
    \label{eq:kpviae}.
\end{align}

When the system is fully symmetric under permutations, all multivariate moments and cumulants of $k$-th order are the same, and equal to the symmetrised quantities $m_k = \langle T_1 T_2 \ldots T_k \rangle$ and $\kappa_k = \llangle T_1 T_2 \ldots T_k \rrangle$. In this case of full statistical equivalence between the particles, the relation between the symmetric multivariate moments $m_k$ and cumulants $\kappa_k$ are exactly the same as between the regular univariate moments $\langle n^k \rangle$ and cumulants $\llangle n^k \rrangle$. Hence the standard combinatorial formula Eq.~\eqref{eq:nkcumviank} applies
\begin{align}
    \kappa_k = \sum_{j=1}^k (j-1)! (-1)^{j-1} B_{kj}
    \left( m_1 , m_2 , \ldots , m_{k-j+1} \right) \, .
    \label{eq:kappaGeneralsupp}
\end{align}  
where $B_{kj}$ are the partial Bell polynomials. For reference, we quote the explicit relations for the values of $k$ involved in the experiment,
\begin{subequations}
\label{eq:kpviamp}
\begin{align}
    \kappa_1 & = m_1 \\
    \kappa_2 & = -m_1^2 +m_2 \\
    \kappa_3 & = 2 \, m_1^3 -3 \, m_2 \, m_1 +m_3 \\
    \kappa_4 & = -6 \, m_1^4 +12 \, m_1^2 \, m_2 -4 \, m_1 \, m_3 -3 m_2^2 +m_4 \\
    \kappa_5 & = 24 \, m_1^5 -60 \, m_1^3 \, m_2 -5 \, m_1 \, m_4 -10\, m_2 \, m_3 +20 \,  m_1^2 \, m_3 +30 \, m_1\, m_2^2 +m_5
\end{align}
\end{subequations}
Together with Eq.~\eqref{eq:mpviapSupp}, these relations are used to compute symmetrised multivariate cumulants form the experimental counting statistics $p_n=P_{(N-n,\,n)}$.

We note that rescaling to spin variables, $s_i=2\, T_i-1$, transforms the multivariate cumulants as  $\llangle s_j \rrangle=2  \llangle T_j \rrangle -1$ and  $\llangle s_1 \ldots s_k \rrangle = 2^k \llangle T_1 \ldots T_k \rrangle $ for $k\ge 2$,
hence $\kappa_{k>1}$ can also be interpreted as irreducible spin correlation functions.

\subsection{Comparison of different types of cumulants}

One can also define factorial cumulants $\llangle n^k \rrangle_\text{F}$ \cite{kambly2011factorial,Kitazawa2017} whose generating function is $\ln \langle e^{zn} \rangle = \sum_{k=1}^{\infty} \llangle n^k \rrangle_\text{F} (z-1)^k/k!$ (expansion of the univariate cumulant generating function around $z=1$).
The corresponding factorial moments, $\langle (n)_k \rangle$, are simply averages of the falling factorial \cite{Kitazawa2017}.
The relations between these factorial cumulants $\llangle n^k \rrangle_\text{F}$ and the factorial moments $\langle (n)_k \rangle = \langle n!/(n-k)! \rangle$ are the same as between univariate cumulants and univariate moments in Eq.~\eqref{eq:nkviankcum} and \eqref{eq:nkcumviank}. 
The advantage of factorial cumulants is evident in a Poisson distribution, where all $\llangle n^k \rrangle_\text{F} = 0$ for $k>1$.

We note that the symmetrised multivariate cumulant $\kappa_k$ is \emph{not} the factorial cumulant $\llangle n^k \rrangle_\text{F}$ of the FCS, because of the factor $(N)_k$ in the relation \eqref{eq:vaifactorialcumulant} between the symmetrised multivariate moment $m_k$ and the factorial moment $\langle (n)_k \rangle$.

The difference between regular univariate cumulants $\llangle n^k \rrangle$, factorial cumulants $\llangle n^k \rrangle_\text{F}$ and symmetrised multivariate cumulants $\kappa_k$ can also be understood in terms of reference probability distributions. 
Namely, $\llangle n^k \rrangle=0$, $\llangle n^k \rrangle_\text{F}=0$ and $\kappa_k=0$ (for $k\ge 2$) are true for Gaussian (in continuous $n$ limit), Poisson and binomial distributions, respectively.

\clearpage
\section{ANALYSIS OF 2e/2e PARTITIONING DATA}
\label{supp:2e2epartitioning}

In Fig.~\ref{figure2}b, four electrons are sent such that the electrons are placed in pairs in two adjacent SAW minima (configuration denoted as 2e/2e and $N=4$). 
This situation is used to explore the case of a less symmetric configuration, with non-equivalent same-order cumulants.

\begin{figure}[t]
    \includegraphics[scale=0.9]{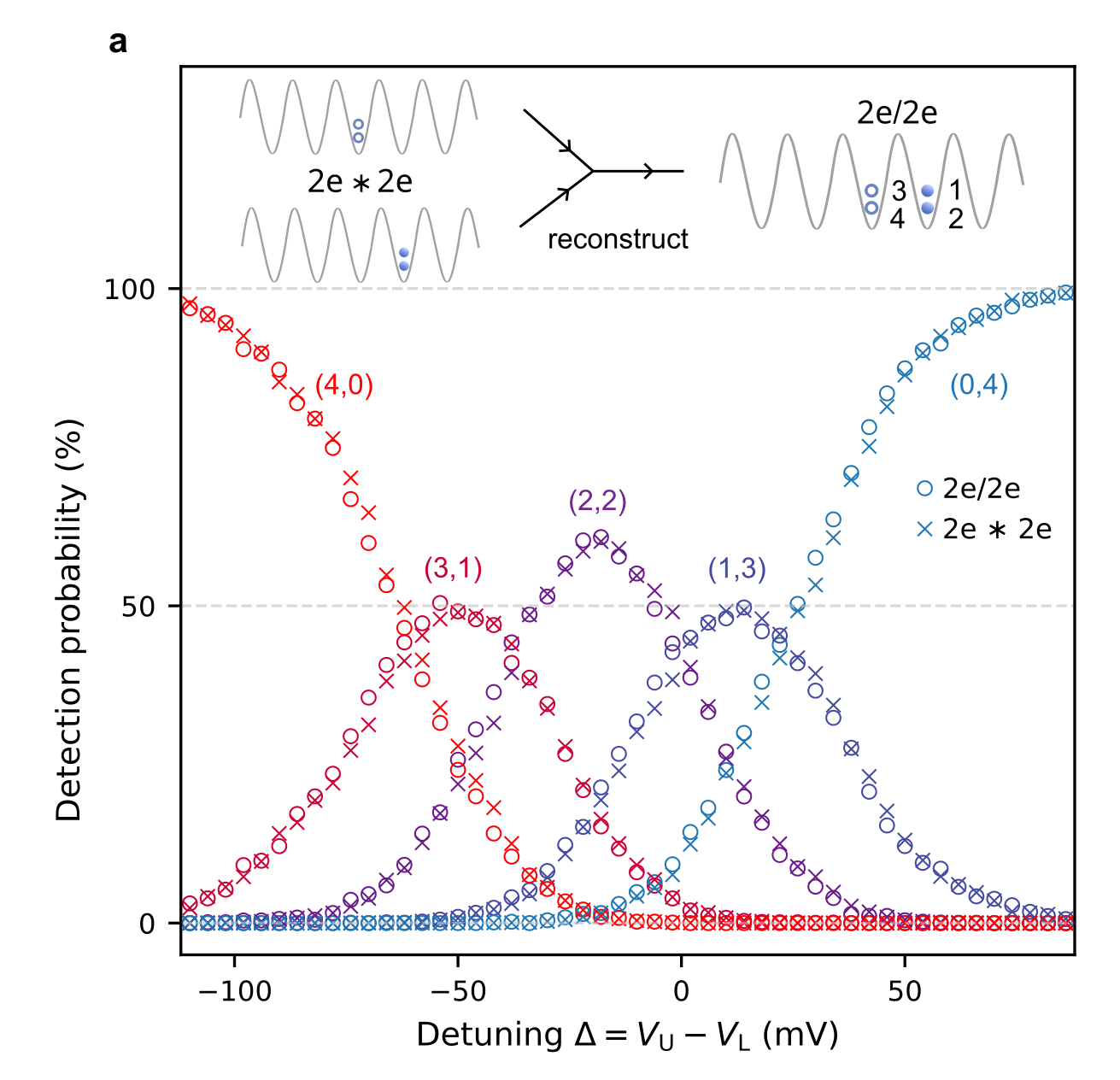}
    \caption{\textbf{Experimental reconstruction of 2e/2e $(N=4)$ from 2e $(N=2)$.} Detection probabilities for the partitioning of four electrons in a configuration with two electrons in each of two adjacent minima (2e/2e) compared to its reconstruction calculated from the detection probabilities measured from the partitioning of two interacting electrons in a single minimum (2e) using Eq.~\eqref{eq:BT}. Each data point is obtained from 3,000 single-shot measurements.}
    \label{figure_S8}
\end{figure}

Similar to the case of one electron per minimum analysed in Fig.~\ref{figure2}a, we can check from the counting statistics that electrons in different SAW minima remain uncorrelated. 
This is done by reconstructing the 2e/2e partitioning probabilities ($N=4$) from the 2e partitioning data where only one minimum is filled with two electrons ($N=2$), using the convolution of corresponding probability distributions
\begin{equation}
\begin{split}
    P_{(4,0)} &=P_{(2,0)}^{2} \\
    P_{(3,1)} &=2P_{(1,1)}P_{(2,0)} \\
    P_{(2,2)} &=P_{(1,1)} ^2+P_{(2,0)}P_{(0,2)} \\
    P_{(1,3)} &=2P_{(1,1)}P_{(0,2)} \\
    P_{(0,4)} &=P_{(0,2)}^{2}. 
\end{split}
\label{eq:BT}
\end{equation}
The measured and reconstructed probabilities are shown in Fig.~\ref{figure_S8} and are in good agreement, confirming the absence of inter-SAW-minima correlations. 

Further, we use this 2e/2e case (with interactions restricted to the electron pairs occupying the same minimum) to illustrate how higher-order multivariate cumulants can be calculated in case of partial permutational symmetry. 
The general formulas~\cite{Gardiner1986HandbookOS} expressing the relations between the multivariate moments $\langle T^{i_1}_{j_1} T^{i_2}_{j_2} \ldots T^{i_l}_{j_l}\rangle$ and the multivariate cumulants can be summarised as follows~\cite{wiedemann2014}
\begin{align}
    \llangle T_{j_1}^{i_1} \dots T_{j_l}^{i_l} \rrangle = \sum_{\lbrace {\cal P}_l\rbrace}
	\left( \vert {\cal P}_l\ \vert - 1\right)!\, \left( -1\right)^{ \vert {\cal P}_l\ \vert - 1}
	\sum_{B \in  {\cal P}_l} \Big\langle \prod_{r\in B} T^{i_r}_{j_r} \Big\rangle\, ,
    \label{eq:multicumviamom}
\end{align}
where ${\lbrace{\cal P}_l\rbrace}$ denotes the list of all partitions of a set $\lbrace 1, 2, \dots, l \rbrace$ for a total of $l$ variables. 
$B\in{\cal P}_l$ is a block in a partition ${\cal P}_l$, and $\vert{\cal P}_l\ \vert$ counts the number of blocks in that partition. 
In our case, we only consider cumulants $\llangle T_{j_1} T_{j_2} \ldots T_{j_l} \rrangle$ as the variables are idempotent ($T_{j}^{2}=T_{j}$).

While general formulas are always valid, they involve too many different cumulants $\llangle T_{j_1} T_{j_2} \ldots T_{j_l} \rrangle$ for all to be determined by the partitioning data. 
However, symmetry considerations may lead to significant reduction in the number of non-equivalent multivariate functions. 
For complete permutational symmetry (when all electrons are statistically indistinguishable, e.g., all placed either in different or in the same SAW minimum), the cumulants $\kappa_k$ are completely determined by counting statistics $p_n$, as described in Methods and Supplementary Note \ref{supp:cumulantsGeneral}. 
In particular, Eq.~\eqref{eq:multicumviamom} leads to Eq.~\eqref{eq:kappaGeneralsupp} with Eq.~\eqref{eq:vaifactorialcumulant} in this case. 
For systems that are not fully symmetric under permutations, not all multivariate cumulants of $k$-th order are equal in general and Eq.~\eqref{eq:kappaGeneralsupp} for calculating the symmetrised cumulants $\kappa_k$ from the symmetrised moments $m_k$ is no longer applicable (however, the cumulants $\kappa_k$ are always defined as in Eq.~\eqref{eq:kpviae}). 

Here we examine a specific case with partial symmetry which follows from considering the particular placement of the electrons in the SAW minima. 
In the present example of 2e/2e electrons, the interaction strength is different between the electrons in the same SAW minimum and the electrons in separate minima. 
We index electrons such that \{1,2\} corresponds to the pair in one minimum (denoted with $\fcircle$) and \{3,4\} to the other pair (denoted with $\hcircle$). 
There are only two non-equivalent second-order multivariate cumulants, one for electrons in the same minimum $\llangle T_1 T_2 \rrangle = \llangle T_3 T_4 \rrangle = \llangle \fcircle \, \fcircle \rrangle$ and one for electrons in different minima $\llangle T_1 T_3 \rrangle = \llangle T_1 T_4 \rrangle = \llangle T_2 T_3 \rrangle = \llangle T_2 T_4 \rrangle = \llangle \fcircle \, \hcircle \rrangle$. 
Hence, following the definition \eqref{eq:kpviae}, $\kappa_2$ is expressed as $\kappa_2=\left( 2\llangle\fcircle\,\fcircle\rrangle + 4\llangle\fcircle\,\hcircle\rrangle \right)/6$. 
The third-order multivariate cumulants $\llangle T_a T_b T_c \rrangle = \llangle\fcircle\,\fcircle\,\hcircle\rrangle = \llangle\fcircle\,\hcircle\,\hcircle\rrangle = \kappa_3$ are all equal, as the subsystem remains physically equivalent for selection of any subset of three electrons (two electrons in the same minimum and one in another). 
Therefore, symmetrised $\kappa_{3}$ can be used for any third-order cumulant in Eq.~\eqref{eq:multicumviamom}. 

Using the above partial symmetry conditions and  Eq.~\eqref{eq:mpviapSupp} and \eqref{eq:multicumviamom}, we express partitioning probabilities $p_n$ via multivariate cumulants (Table~\ref{tab:2e2ecumulants}). 
With two different second-order cumulants, the system is undetermined. 
However, based on the reconstruction of 2e/2e ($N=4$) probability distribution from 2e ($N=2$) partitioning data, we have confirmed that there is no inter-SAW-minima correlation, $\llangle\fcircle\,\hcircle\rrangle=0$. 
This condition allows us to compute the cumulants using the expressions in Table~\ref{tab:2e2ecumulants} by setting $\llangle\fcircle\,\hcircle\rrangle=0$. In this case $\kappa_2=\llangle\fcircle\,\fcircle\rrangle/3$. 
The symmetrised cumulants $\kappa_2$, $\kappa_3$ and $\kappa_4$ computed in this way are depicted in Fig.~\ref{figure2}e of the main text. 
One can see that $\kappa_3$ and $\kappa_4$ are close to zero, demonstrating the absence of higher than second-order correlations.

Alternatively, we could have assumed that $\kappa_4=0$ and then computed $\llangle\fcircle\,\fcircle\rrangle$ and $\llangle\fcircle\,\hcircle\rrangle$ separately (with the expectation that the latter will be close to zero).

\begin{table}
    \centering
    \resizebox{\textwidth}{!}{
    \begin{tabular}{c|c|c|c|c|c|c|c|c|c|c|c|c|c|c|c|c}
        $p_n$ &  $\kappa_{1}$ & $\kappa_{1}^2$ & $\kappa_{1}^3$  & $\kappa_{1}^4$ & $\llangle \fcircle \, \fcircle \rrangle$ & $\llangle \fcircle \, \hcircle \rrangle$ & $\kappa_{1}  \llangle \fcircle \, \fcircle \rrangle$ & $\kappa_{1}  \llangle \fcircle \, \hcircle \rrangle$ & $\kappa_{1}^2  \llangle \fcircle \, \fcircle \rrangle$ & $\kappa_{1}^2  \llangle \fcircle \, \hcircle \rrangle$ & $\llangle \fcircle \, \fcircle \rrangle^2$ & $\llangle \fcircle \, \hcircle \rrangle^2$ & $\kappa_{3}$ & $\kappa_{1} \, \kappa_{3}$ & $\kappa_{4}$ & $1$ \\
        \hline
        $p_0$  & -4  & 6 & -4 & 1 & 2 & 4  & -4 & -8 & 2 & 4 &  1 & 2 & -4  & 4 & 1 & 1\\
        $p_1$  & 4  & -12 & 12 &  -4 & -4 & -8 & 12 & 24 &  -8 & -16 &  -4 & -8 & 12  & -16 & -4 & 0\\
        $p_2$  & 0  & 6 & -12 &  6 & 2 & 4  & -12 & -24 & 12 & 24 &  6 & 12 & -12  & 24 & 6 & 0\\
        $p_3$  & 0  & 0 & 4 &  -4 & 0 & 0 & 4 &  8 & -8 & -16 &  -4 & -8 & 4  & -16 & -4 & 0\\
        $p_4$  & 0  & 0 & 0 &  1 & 0 & 0 & 0 & 0 &  2 & 4 & 1 & 2 &  0 & 4  & 1 & 0 
    \end{tabular}
    }
    \caption{\textbf{Probabilities and cumulants for the 2e/2e case ($N=4$).} 
    Coefficients for expressing the probabilities $p_n=P_{(N-n,\,n)}$ from the multivariate cumulants, taking into account the symmetry but without assumption about the strength of the intra-minimum versus inter-minimum interaction. $p_n$ is the sum of the terms in the top row with each term multiplied by its corresponding coefficient.}
    \label{tab:2e2ecumulants}
\end{table}